%% file: susynlonll.tex
\documentclass{cmspaper}
\usepackage{graphicx}
\usepackage{rotate}
\usepackage{rotating}
\usepackage{relsize}
\usepackage{lineno}
\usepackage{slashed}
\usepackage{slashbox}
\usepackage{url}
\usepackage{amsmath}
\usepackage{longtable}
\usepackage{epstopdf}

\newcommand{\lhc}{{LHC}}
\newcommand{\atlas}{{ATLAS}}
\newcommand{\cms}{{CMS}}
\newcommand{\cteq}{{\tt CTEQ6.6}}
\newcommand{\mstw}{{\tt MSTW2008}}
\newcommand{\gl}{\tilde{g}}
\newcommand{\sq}{\tilde{q}}
\newcommand{\sqb}{\tilde{q}^*}
\newcommand{\st}{\tilde{t}}
\newcommand{\stb}{\tilde{t}^*}
\newcommand{\sbot}{\tilde{b}}
\newcommand{\sbotb}{\tilde{b}^*}
\newcommand{\si}{\sigma}
\newcommand{\NLL}{\mathrm{NLL}}

\newcommand{\NLO}{\mathrm{NLO}}
\newcommand{\rhohat}{\hat{\rho}}
\newcommand{\shat}{\hat{s}}

\newcommand{\pdfctequp}{\ensuremath{\mathrm{PDF}_{\mathrm{CTEQ, up}}}}
\newcommand{\pdfcteqdown}{\ensuremath{\mathrm{PDF}_{\mathrm{CTEQ, down}}}}
\newcommand{\scactequp}{\ensuremath{\mu_{\mathrm{CTEQ, up}}}}
\newcommand{\scacteqdown}{\ensuremath{\mu_{\mathrm{CTEQ, down}}}}
\newcommand{\pdfmstwup}{\ensuremath{\mathrm{PDF}_{\mathrm{MSTW, up}}}}
\newcommand{\pdfmstwdown}{\ensuremath{\mathrm{PDF}_{\mathrm{MSTW, down}}}}
\newcommand{\scamstwup}{\ensuremath{\mu_{\mathrm{MSTW, up}}}}
\newcommand{\scamstwdown}{\ensuremath{\mu_{\mathrm{MSTW, down}}}}
\newcommand{\alphasup}{\ensuremath{\alpha_{\mathrm{S, up}}}}
\newcommand{\alphasdown}{\ensuremath{\alpha_{\mathrm{S, down}}}}

\newcommand{\alphas}{\ensuremath{\alpha_{\mathrm{s}}}}
\newcommand{\cteqnom}{\ensuremath{\mathrm{CTEQ}_{\mathrm{nom}}}}
\newcommand{\mstwnom}{\ensuremath{\mathrm{MSTW}_{\mathrm{nom}}}}
\newcommand{\ctequp}{\ensuremath{\mathrm{CTEQ}_{\mathrm{up}}}}
\newcommand{\cteqdown}{\ensuremath{\mathrm{CTEQ}_{\mathrm{down}}}}
\newcommand{\mstwup}{\ensuremath{\mathrm{MSTW}_{\mathrm{up}}}}
\newcommand{\mstwdown}{\ensuremath{\mathrm{MSTW}_{\mathrm{down}}}}

\def\nn{\nonumber}

\begin{document}

%
\begin{titlepage}

\flushright{CERN-PH-TH-2014-137}
\vspace{-5mm}
\flushright{MS-TP-14-25}
\vspace{-5mm}
\flushright{TTK-14-13}
\vspace{-1mm}
\title{Squark and gluino production cross sections in $pp$ collisions at $\sqrt{s} = 13, 14, 33$ and $100$ TeV}
 \begin{Authlist}
Christoph Borschensky
 \Instfoot{unims}{ Institut f\"ur Theoretische Physik, Westf\"alische Wilhelms-Universit\"at M\"unster, D-48149 M\"unster, Germany}
 Michael Kr\"amer
\Instfoot{rwth}{ Institute for Theoretical Particle Physics and Cosmology, RWTH Aachen University, D-52056 Aachen, Germany}
 Anna Kulesza
 \Instfoot{unims}{ Institut f\"ur Theoretische Physik, Westf\"alische Wilhelms-Universit\"at M\"unster, D-48149 M\"unster, Germany}
Michelangelo Mangano
 \Instfoot{cern}{European Organization for Nuclear Research, CERN, Switzerland}
Sanjay Padhi
 \Instfoot{ucsd}{University of California, San Diego, USA}
Tilman Plehn
 \Instfoot{heidelberg}{Institut f\"ur Theoretische Physik, Universit\"at Heidelberg, Germany}
Xavier Portell 
 \Instfoot{cern}{European Organization for Nuclear Research, CERN, Switzerland}
 \end{Authlist}

\begin{abstract}
We present state-of-the-art cross section predictions for the production of supersymmetric squarks and gluinos at the upcoming LHC run with a centre-of-mass energy of $\sqrt{s} = 13$ and $14$~TeV, and at potential future $pp$ colliders operating at $\sqrt{s} = 33$ and $100$~TeV. The results are based on calculations which include the resummation of soft-gluon emission at next-to-leading logarithmic accuracy, matched to next-to-leading order supersymmetric QCD corrections. Furthermore, we provide an estimate of the theoretical uncertainty due to the variation of the renormalisation and factorisation scales and the parton distribution functions. 
\end{abstract}
\end{titlepage}

\clearpage

\input{intro}

\input{highOrder}

\input{theoryuncert}

\input{coloredStates}
\input{summary}

\section*{Acknowledgements}
We would like to thank W.\ Beenakker, S.\ Brensing-Thewes, R.\ H\"opker, M.\ Klasen, E.\ Laenen, L.\ Motyka,  I.\ Niessen, M.\ Spira and P.M.\ Zerwas for a fruitful collaboration 
on SUSY cross section calculations. In addition, we would like to thank J.\ Rojo for discussions on PDF developments.
This work has been supported by the Helmholtz
Alliance ``Physics at the Terascale'', the DFG SFB/TR9 ``Computational
Particle Physics'', the Foundation for Fundamental Research of Matter (FOM), the Netherlands Organisation for Scientific Research (NWO), 
the  Polish National Science Centre grant, project number DEC-2011/01/B/ST2/03643, and the ERC grant 291377 ``LHCtheory''.  MK thanks the
CERN TH unit for hospitality. SP acknowledges support from the DOE under the grant DOE -FG02-90ER40546.


\end{document}

%% file: intro.tex
\section{Introduction}
\label{sec:intro}
The search for supersymmetry (SUSY) is a central activity of the \lhc\ physics programme. To date, a variety of experimental searches have been performed at the collision energies of $7$ and $8$ TeV and a broad range of possible final states has been examined~\cite{ATLAS_wiki, CMS_wiki}. 

In the framework of the Minimal Supersymmetric extension of the Standard Model (MSSM) with R-parity conservation, SUSY particles are produced in pairs.
At the \lhc, the most copiously produced SUSY particles are expected to be the strongly interacting partners of quarks, the squarks ($\sq$), 
and the partners of gluons, the gluinos ($\gl$).  The dominant squark and gluino pair-production processes are 
\begin{equation}
pp \to \sq\sq, \sq \sqb, \sq\gl, \gl\gl + X \,,
\label{eq:prod}
\end{equation}
together with the charge conjugated processes.  In Eq.~(\ref{eq:prod})
the chiralities of the squarks, $\sq=(\sq_L,\sq_R)$, are suppressed,
and we focus on the production of the partners of the $(u,d,c,s,b)$
quarks which we assume to be mass-degenerate. The production of the
SUSY partners of top quarks, the stops $(\st)$, and, when appropriate,
the partners of bottom quarks, the sbottoms $(\sbot)$, has to be
considered separately due to parton distribution function (PDF) effects and potentially large mixing
affecting the mass splittings. In this case, we explicitly specify the
different mass states in the pair-production processes,
\begin{equation}
pp \to \st_i \stb_i, \sbot_i \sbotb_i + X \qquad\qquad i=1,2\,,
\label{eq:stsb_prod}
\end{equation} 
where $i=1,2$ corresponds to the lighter and heavier states, respectively.


Given the importance of SUSY searches at the \lhc, accurate knowledge of theoretical predictions for the cross sections is required. Starting from mid-2011, ATLAS and CMS analyses have been based on resummed results at the next-to-leading logarithmic (NLL) accuracy matched to next-to-leading order (NLO) predictions, referred to as NLO+NLL in the rest of this paper. 

With minimal assumptions on SUSY productions and decays, the interpretation of the current \lhc\ data with $\sqrt{s} = 7$ and $8$ TeV leads to the mass bound for the gluino and light squarks as $m_{\tilde{g}} = m_{\tilde{q}} > 1.7 $ TeV, or $m_{\tilde{g}} > 1.4$ TeV with a decoupled squark sector, and $m_{\tilde{q}} > 850 $ GeV with the other decoupled particles, based on the ATLAS/CMS studies. Pair productions of SUSY third generation lighter states also have a mass bound above $m_{\tilde{t}, \tilde{b}} > 750$ GeV. This paper provides a reference for the evaluation of SUSY squark and gluino production cross sections and their theoretical uncertainties for the extended range of superpartner masses within the reach of the upcoming LHC runs at $\sqrt{s}=13$ and $14$~TeV, and for future high-energy hadron colliders at $\sqrt{s}=33$ and  $100$~TeV.
The paper follows~\cite{Kramer:2012bx} in which results for $\sqrt{s}=7$~TeV were presented. The detailed cross section values for the
relevant processes and SUSY models considered by the
experiments, as well as the results for lower LHC centre-of-mass
energies, are collected at the SUSY cross section working group 
web page~\cite{combined7TeV}.


The next section briefly describes the current state-of-the-art higher
order calculations for squark and gluino hadroproduction, followed by the prescription used for the
treatment of theoretical uncertainties in section~\ref{sec:uncert}. In
section~\ref{sec:xsectplots} the production
cross sections are presented, and a summary of the results and of the 
future prospects is given in section~\ref{sec:summary}.

%% file: highOrder.tex

\section{Higher order calculations -- NLO+NLL}
\label{sec:highorder}

The dependence of hadron collider observables on the renormalisation
and factorisation scales is an artifact of perturbation theory and is
generically 
reduced as higher-order perturbative contributions are included. 
Assuming that there is no systematic shift of an
observable from order to order in perturbation theory, for example due
to the appearance of new production channels, the range of rates
covered by the scale
dependence at a given loop order should include
the true prediction of this rate. 
The scale dependence therefore provides a lower limit on the theory
uncertainty of a QCD prediction, which becomes smaller as higher-order
SUSY-QCD corrections are included.
To estimate the scale uncertainty in this study we 
vary simultaneously factorisation and renormalisation scales, within a
range of 0.5 to 2 times the reference central scale $\mu$, where 
$\mu$ is the average of the two sparticle masses in the final state.

The corrections often increase the size of
the cross section with respect to the leading-order prediction
\cite{Kane:1982hw, Harrison:1982yi,Dawson:1983fw} if the
renormalisation and factorisation scales are chosen close to the
average mass of the produced SUSY particles. As a result, the SUSY-QCD
corrections have a substantial impact on the determination of mass
exclusion limits and would lead to a significant reduction of uncertainties
on SUSY mass or parameter values in the case of discovery, see e.g.\ \cite{Beenakker:2011dk}.
The
processes listed in Eqs.~(\ref{eq:prod}) and (\ref{eq:stsb_prod})
have been known for quite some time at NLO  in
SUSY-QCD~\cite{Beenakker:1994an,Beenakker:1995fp, Beenakker:1996ch,
  Beenakker:1997ut}. Note that SUSY-QCD corrections can be split into
two parts. First, there are QCD corrections induced by gluon or quark
radiation and by gluon loops, which follow essentially the
same pattern as, for example, top pair production. Second, there are virtual 
diagrams which involve squark and gluino loops, and which are independent of the real emission corrections. 
For heavy squarks and gluinos, the virtual SUSY loops are numerically sub-leading, albeit
challenging to compute. 
This is mainly due to a large number of Feynman diagrams 
 with different mass scales contributing to the overall cross section.
For stop pair production, where neither light-flavour squarks nor gluinos
appear in the tree-level diagrams, the virtual SUSY contributions can be easily
decoupled~\cite{Beenakker:1997ut}. The only part that requires some
attention is the appropriate treatment of the counter term and the
running of the strong coupling constant. This decoupling limit is
implemented in {\sc Prospino2}~\cite{prospino2}. For light-flavour squark and gluino
production this decoupling would only be consistent if applied to the
leading order as well as NLO contributions. This is usually not required, 
unless we choose specific simplified models.

Given the expected squark flavour structure in the MSSM, most
numerical implementations, including {\sc Prospino2}, make assumptions
about the squark mass spectrum. The left-handed and right-handed
squarks of the five light flavours are assumed to be mass
degenerate. Only the two stop masses are kept separate in the NLO
computations of light-flavour production
rates~\cite{Beenakker:1994an,Beenakker:1995fp, Beenakker:1996ch}. In
the {\sc Prospino2}~\cite{prospino2} implementation, this degeneracy is not assumed for
the leading-order results. However, the approximate NLO rates are
computed from the exact leading order cross sections times the mass
degenerate $K$-factors, i.e.\ the ratio of NLO and LO cross section for
mass degenerate squarks. For the pair production of third-generation
squarks the four light squark flavours are assumed to be mass degenerate, while the
third generation masses are kept
separate~\cite{Beenakker:1997ut}. This approximation can for example
be tested using {\sc MadGolem}~\cite{Binoth:2011xi}, an automatised NLO
tool linked to {\sc MadGraph4}~\cite{Alwall:2007st}, or other 
recent NLO calculations that keep all squark masses
separate~\cite{Hollik:2012rc, Hollik:2013xwa, GoncalvesNetto:2012yt, Gavin:2013kga}. 
It is also important to point out here that in {\sc Prospino2} the pair production of
third-generation squarks is available as an individual
processes. However, sbottom pairs are included in the implicit sum of
light-flavour squarks because there is no perfect separation of bottom
and light-flavour decay jets.

When summing the squark and gluino production rates including
next-to-leading order corrections it is crucial to avoid double
counting of processes. For example, squark pair production includes
$\mathcal{O}(\alphas^3)$ processes of the kind $qg \to \sq \sq^*q$. 
The same final state can be produced in $\sq \gl$ production when
the on-shell gluino decays into an anti-squark and a quark. The {\sc
  Prospino} scheme for the separation and subtraction of on-shell
divergences from the $\sq \sq^*$ process uniquely ensures a consistent
and point-by-point separation over the entire phase space, see also \cite{Gavin:2013kga}. For a
finite particle mass this scheme has recently been adopted by {\sc MC\@@NLO}~\cite{Weydert:2009vr}
for top quark processes. It is automatised as part of {\sc
  MadGolem}~\cite{Binoth:2011xi}.

A significant part of the NLO QCD corrections can be attributed to the
threshold region, where the partonic centre-of-mass energy is close to
the kinematic production threshold. In this case the NLO
corrections are typically large, with the most significant
contributions coming from soft-gluon emission off the coloured
particles in the initial 
and final state. The contributions due to soft gluon emission can be
consistently taken into 
account to all orders by means of threshold resummation. In this
paper, we discuss results where resummation has been performed at 
next-to-leading logarithmic (NLL) accuracy~\cite{Kulesza:2008jb,Kulesza:2009kq,
Beenakker:2009ha,Beenakker:2010nq,Beenakker:2011fu}. 

The step from NLO to NLO+NLL is achieved by calculating the NLL-resummed partonic cross section 
$\tilde \sigma^{\rm (NLL)}$ and then matching it to the NLO prediction, in order to retain the available information 
on other than soft-gluon contributions. The matching procedure takes the following form
\begin{eqnarray}
\label{eq:matching}
\si^{\rm (NLO+NLL)}_{p p \to kl}\bigl(\rho, \{m^2\},\mu^2\bigr) 
  &=& \si^{\rm (NLO)}_{p p \to kl}\bigl(\rho, \{m^2\},\mu^2\bigr)\nn
          \\[1mm]
   &&  \hspace*{-30mm}+\, \frac{1}{2 \pi i} \sum_{i,j=q,\bar{q},g}\, \int_\mathrm{CT}\,dN\,\rho^{-N}\,
       \tilde f_{i/p}(N+1,\mu^2)\,\tilde f_{j/p}(N+1,\mu^2) \nn\\[0mm]
   && \hspace*{-20mm} \times\,
       \left[\tilde\si^{\rm(NLL)}_{ij\to kl}\bigl(N,\{m^2\},\mu^2\bigr)
             \,-\, \tilde\si^{\rm(NLL)}_{ij\to kl}\bigl(N,\{m^2\},\mu^2\bigr)
       {\left.\right|}_{\scriptscriptstyle({\NLO})}\, \right]\,,
\end{eqnarray}
where the last term in the square brackets denotes the NLL resummed
expression expanded to NLO. The symbol $\{m^2\}$ stands for all masses entering the 
calculations and  $\mu$ is the common factorisation and
renormalisation scale. The resummation is performed in the Mellin moment $N$ space, 
with all Mellin-transformed quantities indicated by a tilde. In particular, the Mellin moments of the 
partonic cross sections are defined as 
\begin{equation}
\label{eq:Mellin}
  \tilde\si_{i j \to kl}\bigl(N, \{m^2\}, \mu^2 \bigr) 
 \equiv \int_0^1 d\rhohat\;\rhohat^{N-1}\;
           \si_{ i j\to kl}\bigl(\rhohat,\{ m^2\}, \mu^2 \bigr) \,.
\end{equation}
The variable $\rhohat \equiv (m_k + m_l)^2/\shat $ measures the closeness to the partonic production threshold 
and is related to the corresponding hadronic variable $\rho = \rhohat x_i x_j$ in Eq.~(\ref{eq:matching}), where $x_i\ (x_j)$ is the usual longitudinal momentum fraction of the incoming proton carried out by the parton $i (j)$. 
The necessary inverse Mellin transform in Eq.~(\ref{eq:matching}) is performed  along the contour ${\rm CT}$ according to the 
so-called ``minimal prescription''~\cite{Catani:1996yz}.
The NLL resummed cross section in
Eq.~(\ref{eq:matching}) reads
\begin{eqnarray}
  \label{eq:NLLres}
  \tilde{\sigma}^{\rm (NLL)} _{ij\rightarrow kl}\bigl(N,\{m^2\},\mu^2\bigr) 
& = &\sum_{I}\,
      \tilde\sigma^{(0)}_{ij\rightarrow
        kl,I}\bigl(N,\{m^2\},\mu^2\bigr)\, \nn  \\[1mm]
   & \times\,& \Delta^{\rm (NLL)}_i (N+1,Q^2,\mu^2)\,\Delta^{\rm (NLL)}_j (N+1,Q^2,\mu^2)\,
     \Delta^{\rm (s, \NLL)}_{ij\rightarrow
       kl,I}\bigl(N+1,Q^2,\mu^2\bigr)\,,
\end{eqnarray}
where the hard scale $Q^2$ is taken as $Q^2 = (m_k + m_l)^2$ and $\tilde{\sigma}^{(0)}_{ij \rightarrow kl, I}$ are the
colour-decomposed leading-order cross sections in Mellin-moment space,
with $I$ labelling the possible colour
structures.  The functions $\Delta^{\rm (NLL)}_{i}$ and $\Delta^{\rm (NLL)}_{j}$ sum
the effects of the (soft-)collinear radiation from the incoming
partons. They are process-independent and do not depend on the colour
structures.  These functions contain both the leading logarithmic as well
as part of the sub-leading logarithmic behaviour. The expressions for
$\Delta^{\rm (NLL)}_{i}$ can be found in the
literature~\cite{Kulesza:2009kq}. In order to perform resummation at NLL accuracy, one also has to take 
into account soft-gluon contributions involving emissions from the final state, depending on the colour structures in which the
final state SUSY particle pairs can be produced. They are summarised by the factor
\begin{equation}
  \Delta_{I}^{\rm (s, \NLL)}\bigl(N,Q^2,\mu^2\bigr) 
  \;=\; \exp\left[\int_{\mu}^{Q/N}\frac{dq}{q}\,\frac{\alphas(q)}{\pi}
                 \,D_{I} \,\right]\,.
\label{eq:2}
\end{equation}
The one-loop coefficients $D_{I}$ follow from the threshold limit of
the one-loop soft anomalous-dimension
matrix and can be found in~\cite{Kulesza:2009kq,Beenakker:2009ha}. 

The analytic results for the NLL part of the cross sections have been implemented into a numerical code. 
The results of this code, added to the NLO results obtained from {\sc Prospino2}, correspond to the matched NLO+NLL cross sections. 
Their central values, the scale uncertainty and the 68\% C.L. PDF and $\alphas$ uncertainties obtained using \cteq~\cite{cteq66} and \mstw~\cite{mstw08} PDFs have been tabulated for the squark and gluino production processes of interest in the range of input masses appropriate for the analyses.\footnote{\protect Note that we use NLO PDFs with the NLO+NLL matched cross section calculation. The reduction
of the factorisation scale dependence observed in the NLO+NLL predictions is a result of a better
compensation between the scale dependence of the NLO evolution of the PDF and the short
distance cross section, and does not depend on whether PDFs are fitted using NLO or NLO+NLL
theory, see for example Ref.~\cite{Sterman:2000pu}. In general, one can apply NLL
threshold resummation with
NLO PDFs to processes like heavy SUSY particle production for which the summation of logarithms
is more important than for the input data to the NLO fits. However, it would be interesting to systematically study the difference  
between NLO and NLO+NLL input to global PDF determinations for SUSY
particle production at the LHC.}
Together with a fast interpolation code, the tabulated values constitute the {\sc NLL-fast} 
numerical package~\cite{Beenakker:2011fu,nllfast}.

In this paper we present results based on NLO+NLL calculations which can be obtained with the {\sc NLL-fast}  package. Results for squark and gluino production at next-to-next-to-leading-logarithmic (NNLL)  level  in collisions at 8 TeV have been recently presented in~\cite{Beenakker:2014sma}  and a numerical code is in development. Additional NNLL results are available for selected processes  such as 
stop-antistop \cite{Broggio:2013uba} and gluino-gluino pair~\cite{Pfoh:2013iia} production. For these processes, approximate next-to-next-to-leading order (NNLO) predictions including the dominant NNLO corrections coming from the resummed cross section at
next-to-next-to-leading-logarithmic (NNLL) level, also exist \cite{Langenfeld:2009eg, Langenfeld:2010vu, Langenfeld:2012ti}. Moreover, a general
formalism has been developed in the framework of effective field
theories which allows for the resummation of soft and Coulomb gluons
in the production of coloured
sparticles~\cite{Beneke:2009rj,Beneke:2010da} and subsequently applied to squark and gluino production at NLL~\cite{Beneke:2010da, Falgari:2012hx} and NNLL accuracy~\cite{Beneke:2013opa}. 
The production of gluino
bound states as well as bound-state effects in gluino pair and
squark-gluino production
has also been studied \cite{Hagiwara:2009hq,Kauth:2009ud,Kauth:2011vg,Kauth:2011bz}, with a recent study at NNLL accuracy~\cite{Kim:2014yaa} concentrating on the stoponium bound states. Finite width effects in the production of squark and gluino pairs have been investigated in~\cite{Falgari:2012sq}.
Furthermore, electroweak corrections to the ${\cal O} (\alphas^2)$ tree-level
processes~\cite{Hollik:2007wf, Hollik:2008yi, Hollik:2008vm, Beccaria:2008mi,
    Mirabella:2009ap, Germer:2010vn, Germer:2011an} and the
electroweak Born production channels of ${\cal O} (\alpha\alphas)$
and ${\cal O} (\alpha^2)$~\cite{Alan:2007rp,Bornhauser:2007bf} are in
general significant for the pair production of SU(2)-doublet squarks
$\tilde{q}_L$ and at large invariant masses, but they are moderate for
inclusive cross sections and will not be included in the results
presented here. 


%% file: theoryuncert.tex
\section{Treatment of cross sections and their associated uncertainties}
\label{sec:uncert}

The cross sections are taken at the
next-to-leading order in the strong coupling constant, including the
resummation of soft gluon emission at the NLL level of accuracy, performed using
the {\sc{NLL-fast}} code. Currently, the code provides predictions for all squark and gluino production processes at $\sqrt{s}=7, 8$ and $13$ TeV. Additionally, results for stop (sbottom) pair production, gluino pair production with decoupled squarks and squark production with decoupled gluinos are available for $\sqrt{s}=13, 14, 33$ and 100 TeV~\cite{nllfast}. For these particular cases, {\sc{NLL-fast}} delivers cross sections for masses spanning
$200$~GeV to $3, 3.5, 6.5, 15$~TeV for squark and gluino production 
and $100$~GeV to $2.5, 2.5, 5, 10$~TeV for direct stop or sbottom pair
production at  $\sqrt{s}=13, 14, 33, 100$ TeV, correspondingly. Further updates will appear soon~\cite{nllfast}.  Following the
convention used in {\sc Prospino2}, in the case of
squarks, which can be more or less degenerate depending on a specific SUSY
scenario, the input mass used is the result of
averaging only the first and second generation squark masses. 
Further details on different scenarios considered to interpret the variety of
experimental searches developed by the \atlas\ and \cms\ collaborations
are described in Section~\ref{sec:special}.


Scenarios have been investigated in which either the squark or gluino mass are set to some
high scale, such that the corresponding sparticles cannot be produced
at the LHC. Defining such a large mass scale is of course to some
extend arbitrary and may have a non-negligible impact on the production of the SUSY particles residing at the TeV scale (e.g. squarks at
high scales can still contribute to the gluino pair production process
via a t-channel exchange). Thus, the calculation implemented in
{\sc{NLL-fast}} assumes that very heavy squarks or gluinos are completely decoupled and do
not interfere with the production processes of the kinematically
accessible particles. 

The uncertainties due to the choice of the renormalisation and
factorisation scales as well as the PDFs are obtained using the
{\sc{NLL-fast}} code. 
In order to combine all these predictions and obtain an overall
uncertainty estimate, the {\sc PDF4LHC} recommendations are followed as closely
as possible, based on the availability of different
calculations. Thus, an envelope of cross section predictions is
defined using the 68\% C.L. ranges of the \cteq~\cite{cteq66}
(including the $\alphas$ uncertainty) and \mstw~\cite{mstw08} PDF
sets, together with the variations of the scales. The nominal cross
section is obtained using the midpoint of the envelope and the
uncertainty assigned is half the full width of the
envelope. If \pdfctequp\ (\pdfcteqdown) and
\scactequp\ (\scacteqdown) are the upward (downward) one sigma
variations of the \cteq\ PDF set, respectively,
\pdfmstwup\ (\pdfmstwdown) and \scamstwup\ (\scamstwdown) are the
corresponding variations for the \mstw\ PDF set and, finally,
\alphasup\ (\alphasdown) is the corresponding up (down) one sigma
uncertainty of the \alphas\ coupling constant, the following
quantities can be calculated:
\begin{subequations}
\begin{align}
\label{eq:variations}
&\ctequp = \sqrt{\pdfctequp^2 + \scactequp^2+\alphasup^2} \,,\\
&\cteqdown = \sqrt{\pdfcteqdown^2 + \scacteqdown^2+\alphasdown^2}\,, \\
&\mstwup = \sqrt{\pdfmstwup^2 + \scamstwup^2} \,,\\
&\mstwdown = \sqrt{\pdfmstwdown^2 + \scamstwdown^2}\,.
\end{align}
\end{subequations}
\noindent The corresponding upper and lower values of the envelope created by this set of numbers and the nominal predictions (\cteqnom\ and \mstwnom) are obtained by:
\begin{subequations}
\begin{align}
\label{eq:UandL}
&\mathrm{U} = \mathrm{max} (\cteqnom + \ctequp, \mstwnom + \mstwup )\,, \\
&\mathrm{L} = \mathrm{min} (\cteqnom - \cteqdown, \mstwnom - \mstwdown )\,,
\end{align}
\end{subequations}
\noindent and the final corresponding cross section ($\sigma$) and its symmetric uncertainty ($\Delta\sigma$) are taken to be: 
\begin{subequations}
\begin{align}
\label{eq:sigmas}
\sigma &= (\mathrm{U}+\mathrm{L})/2 \,,\\
\Delta\sigma &= (\mathrm{U}-\mathrm{L})/2\,.
\end{align}
\end{subequations}
Full compliance with the PDF4LHC recommendations, with the inclusion
of other PDF sets such as {\sc NNPDF}~\cite{nnpdf}, will be
implemented in {\sc NLL-fast} or its successor. We notice that, as discussed in
section~\ref{sec:xsectplots}, the additional contribution to the
systematics coming from \alphas\ uncertainties is negligible.

\subsection{Special cases}
\label{sec:special}
Some SUSY models require special treatment in order to ensure that the
NLO cross sections are correctly computed.  
Given the difficulty
to provide a comprehensive summary of all situations that are
being considered in the interpretation of the LHC data, we only
discuss here few relevant cases, to exemplify the approach followed.

\subsubsection*{Simplified Models}

A variety of simplified models~\cite{Alves:2011wf} are considered by the experiments. In some cases, the gluino, 
sbottom, and stops are decoupled from the rest of the
supersymmetric spectrum.\footnote{While this scenario is possible if we only
consider the TeV scale, it bears some challenges at higher energy
scales. Any kind of renormalisation scale evolution will generate
squark masses at the scale of the gluino mass, but not vice versa. 
Thus,  any discovery of light squarks associated with heavy
gluinos would point to a non-standard underlying
model~\cite{Jaeckel:2011wp}. In spite of all theory prejudice it is
clearly adequate that these regions be experimentally explored.}
In this specific simplified model, only squark-antisquark production
is allowed and this process is flavour-blind, if the masses are
considered degenerate. Since the NLO+NLL calculations consider the
sbottom as degenerate in mass with the squarks of the first and second
generations, the overall cross section has to be rescaled by a
factor of $4/5=0.8$.

In other cases where the gluino is not decoupled, 
squark-gluino and squark-squark productions are feasible, and can be
used as provided by default. The only effect could come from a $b$-quark in the initial state,
which however is strongly suppressed numerically~\cite{Beenakker:2010nq}.
Other types of simplified models decouple not only the third
generation squarks, but in addition all
right-handed squarks.  These scenarios primarily focus on squark decays via charginos or neutralinos. 
The squark mass is then calculated by averaging the non-decoupled
squark masses and the final cross section is scaled by 
a factor of $(4/5)\cdot (1/2)=0.4$.

\subsubsection*{Treatment of 3rd generation squarks}
Direct stop and sbottom production must be treated
differently from the rest of squark families because, for instance,
the $t$-channel gluino-exchange diagrams are suppressed. 
In computations of squark production processes 
involving sbottoms, masses of sbottoms can be considered either degenerate with the rest of other squark flavours, as done in {\sc{NLL-Fast}}, or non-degenerate. In scenarios
in which the production of different squark flavours are present, the squark pair production cross
section is rescaled down to subtract the sbottom contribution and the
corresponding process is computed separately.

At leading order the corresponding partonic cross sections for the production of third-generation squarks 
depend only on their masses, and the results for sbottom and stop of
the same mass are therefore equal. At NLO in SUSY-QCD, additional SUSY
parameters like squark and gluino masses or the stop/sbottom mixing
angle enter. Their numerical impact, however, is very
small~\cite{Beenakker:1997ut, Beenakker:2010nq}. 
A further
difference between stop and sbottom pair production arises from the 
$b\bar{b} \to \tilde{b}\tilde{b}^*$ channel, where the initial-state
bottom quarks 
do allow a $t$-channel gluino-exchange graph that gives rise to extra
contributions. However, as has been demonstrated in Ref.~\cite{Beenakker:2010nq} their
numerical impact on the hadronic cross sections is negligible. Thus,
for all 
practical purposes, the LO and higher-order cross section predictions 
obtained for stop pair production apply also to sbottom pair
production if the 
input parameters, i.e.\ masses and mixing angles, are modified accordingly.

%% file: coloredStates.tex

\newcommand{\myxsectcaption}[3] {NLO+NLL {#1} production cross section
  with {#2} decoupled  as a function of mass at $\sqrt{s}={#3}$ TeV in the wider (upper plot) and narrower (lower plot) mass range. The different styled black (red) lines
  correspond to the cross section and scale uncertainties predicted
  using the \cteq\ (\mstw) PDF set. The yellow (dashed black) band
  corresponds to the total \cteq\ (\mstw) uncertainty, as described in
  the text. The green lines show the final cross section and its total
  uncertainty.}

\newcommand{\nlonllcaption}[1] {NLO+NLL production cross sections for the case of equal degenerate squark and gluino masses as a function of mass at $\sqrt{s}={#1}$ TeV. }

\section{Squark and gluino production at the LHC}
\label{sec:xsectplots}

\begin{figure}[htp]
  \begin{center}
       	\includegraphics[width=12cm]{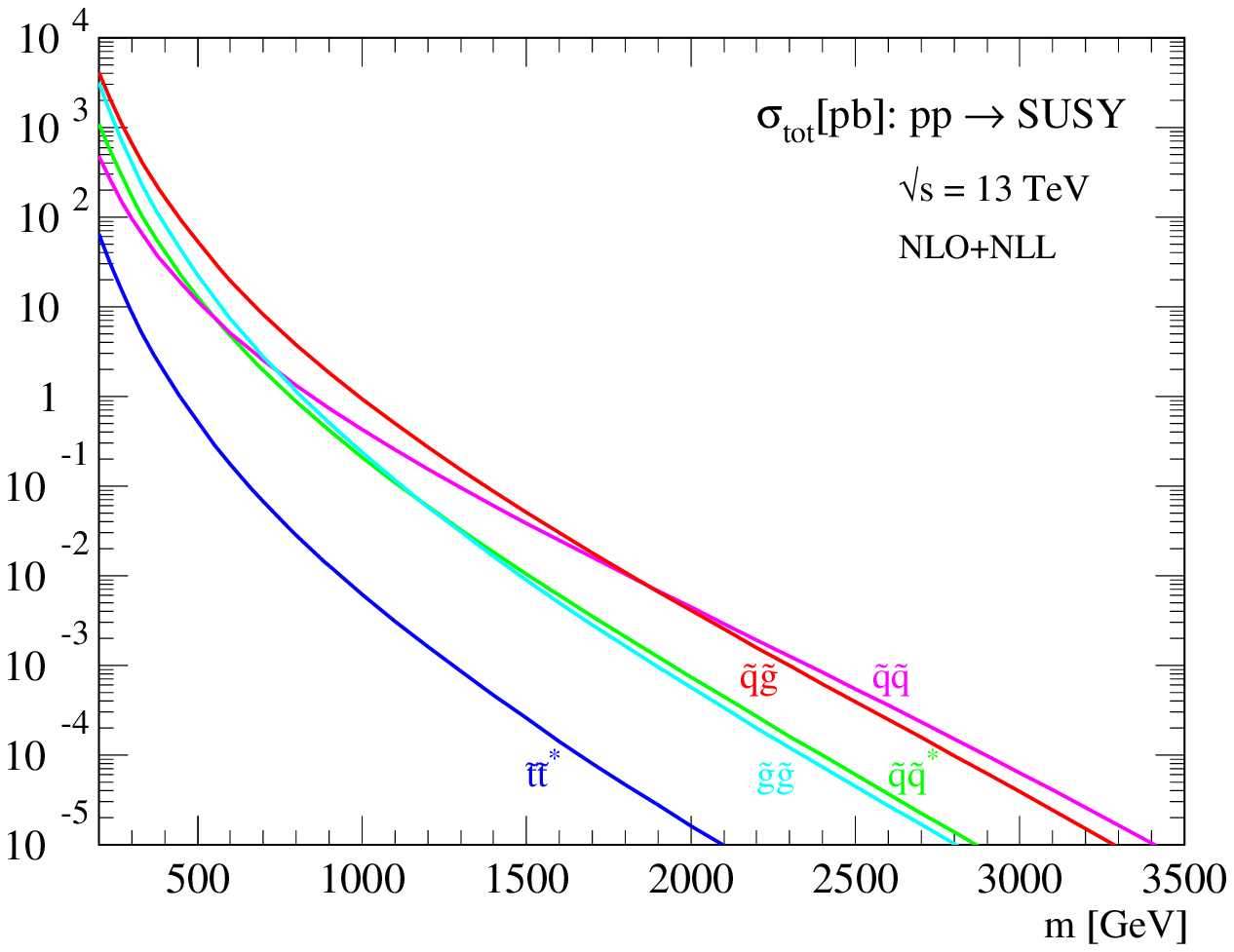}
  \end{center}
  \caption{\nlonllcaption{13}}
  \label{fig:xsect_all_13}
\end{figure}

\begin{figure}[htp]
  \begin{center}
       	\includegraphics[width=12cm]{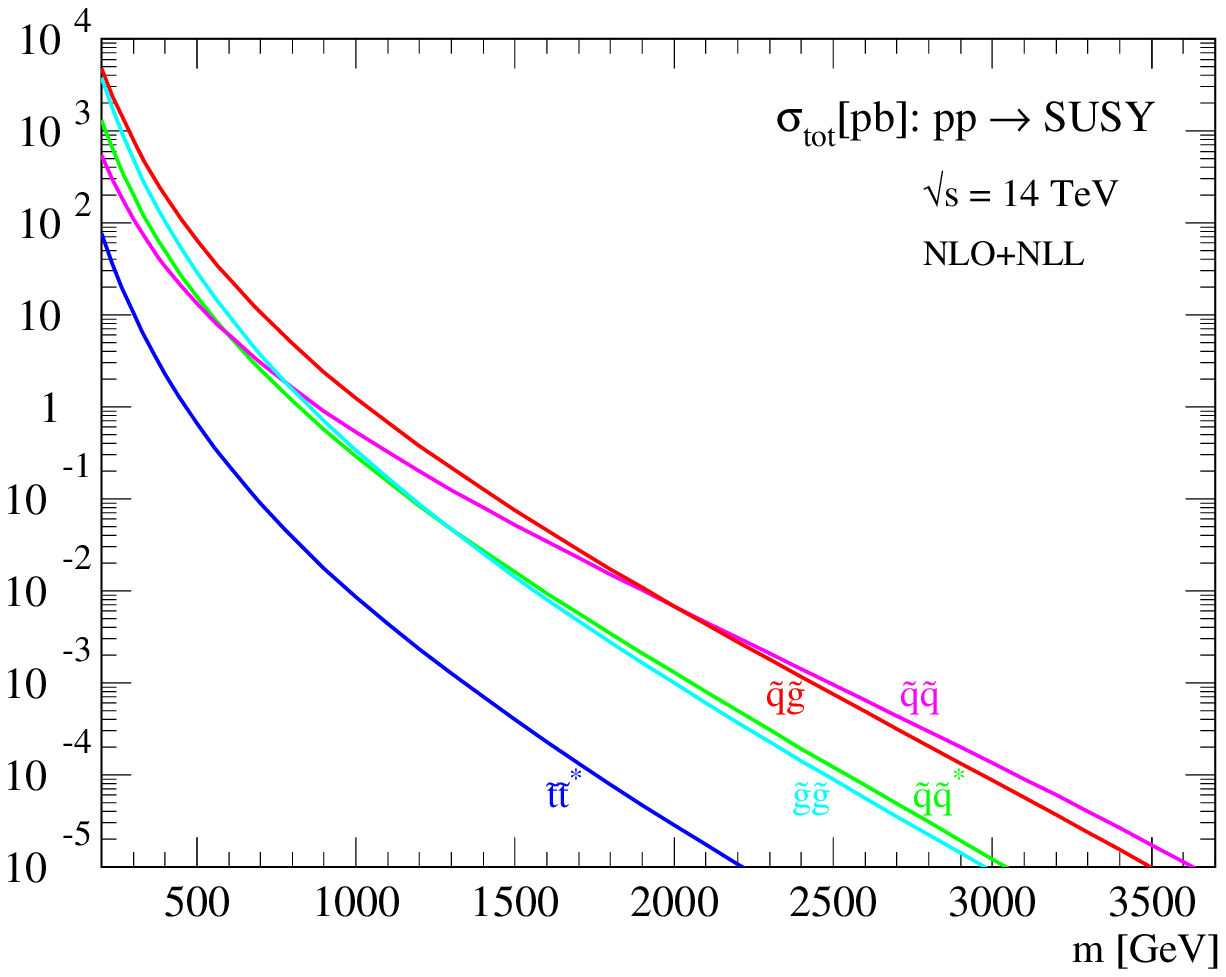}
  \end{center}
  \caption{\nlonllcaption{14}}
  \label{fig:xsect_all_14}
\end{figure}

\begin{figure}[hbp]
  \begin{center}
       	\includegraphics[width=12cm]{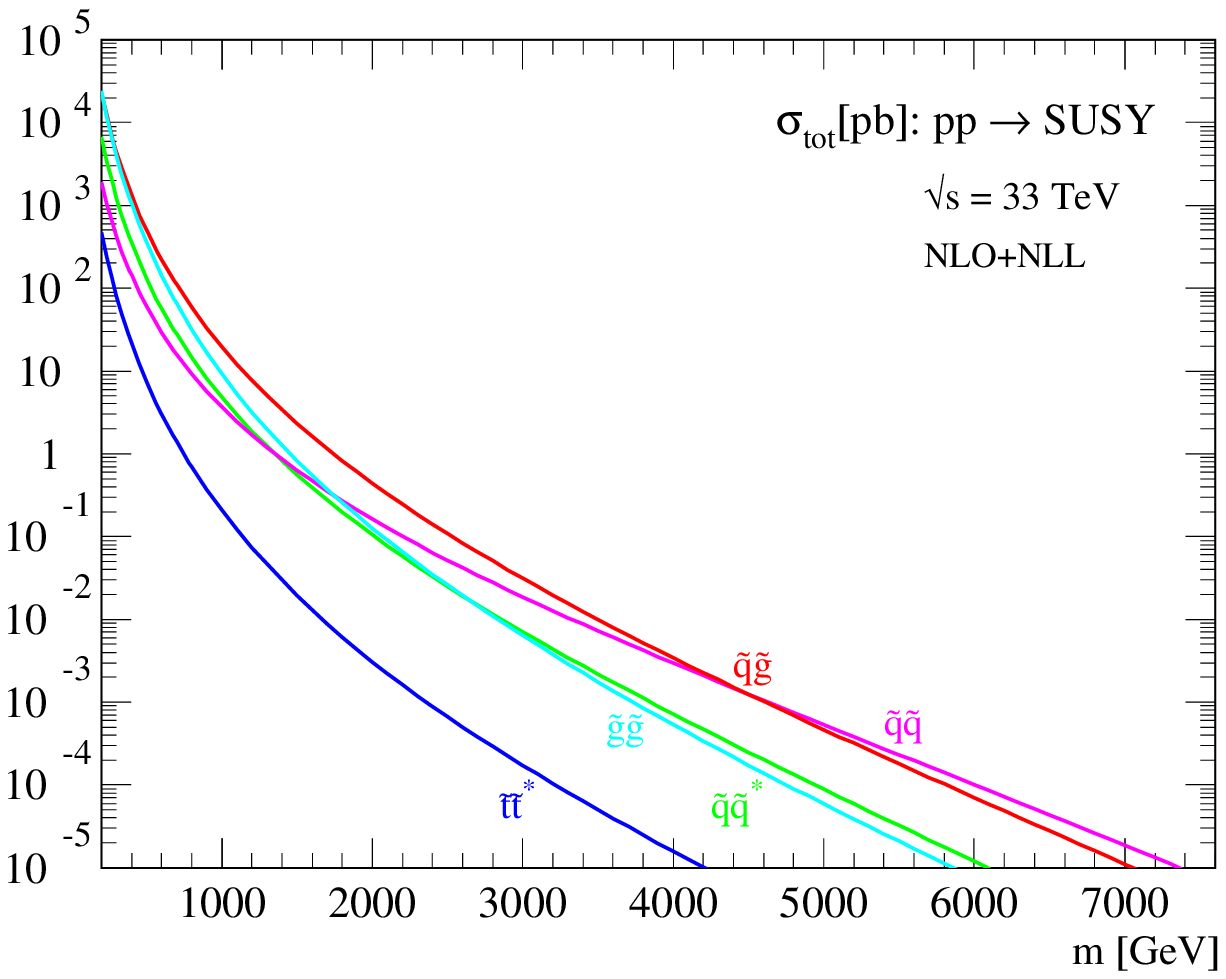}
  \end{center}
  \caption{\nlonllcaption{33}}
  \label{fig:xsect_all_33}
\end{figure}

\begin{figure}[htp]
  \begin{center}
       	\includegraphics[width=12cm]{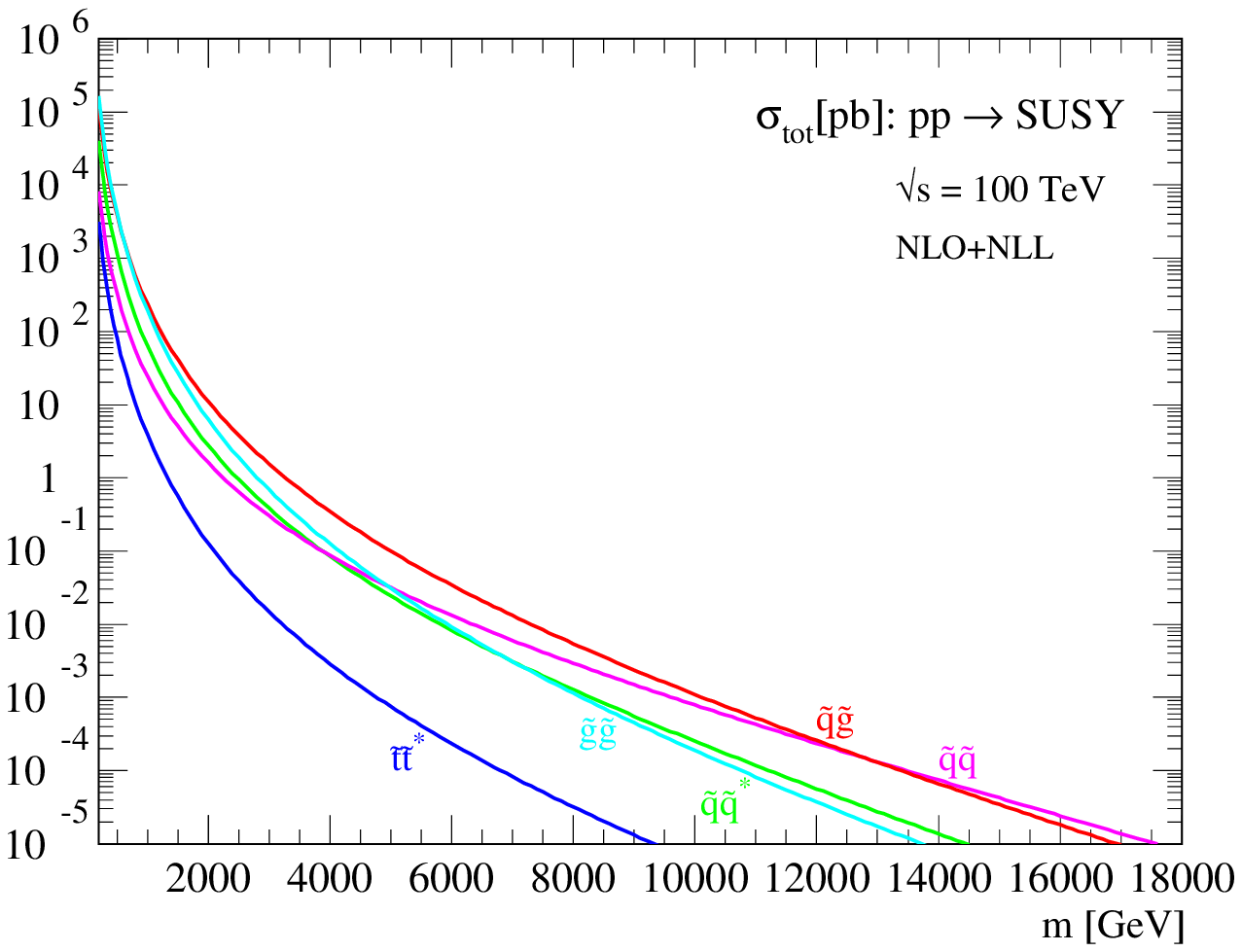}
  \end{center}
  \caption{\nlonllcaption{100}}
  \label{fig:xsect_all_100}
\end{figure}

The production cross sections and associated uncertainties
resulting from the procedure described in the previous section are
discussed here for different processes of interest. First, in
Figures~\ref{fig:xsect_all_13},~\ref{fig:xsect_all_14},~\ref{fig:xsect_all_33}
and~\ref{fig:xsect_all_100} we show the NLO+NLL central predictions
for the various squark and gluino production processes 
for the case of equal squark and gluino masses at collider energies
$\sqrt{s}=13, 14, 33$ and $100$ TeV, respectively. 
Assuming a squark and gluino mass near 2~TeV, we predict inclusive
SUSY cross sections of the order $10^{-2}$, 1 and 20~pb at $\sqrt{s}=13 (14),
33$ and $100$ TeV, respectively. The relative
size of the various production channels depends on the relative size
of squark/gluino masses and collider energy. For small SUSY masses
and/or large collider energies, gluino cross sections are dominant,
while for large SUSY masses and/or
low collider energies the valence quark distributions favour
squark-gluino associate and squark-pair production. 

We furthermore discuss three distinct special cases in some detail:
gluino pair production with decoupled squarks, squark-antisquark pair
production with gluino decoupled and stop/sbottom pair production. The
results shown here are mainly illustrative: tables with cross sections and
systematic uncertainties obtained in other scenarios 
are collected at the SUSY cross section working group web page~\cite{combined7TeV}.

\subsection{Gluino pair production}

\begin{figure}[htbp]
  \begin{center}
       	\includegraphics[width=12cm]{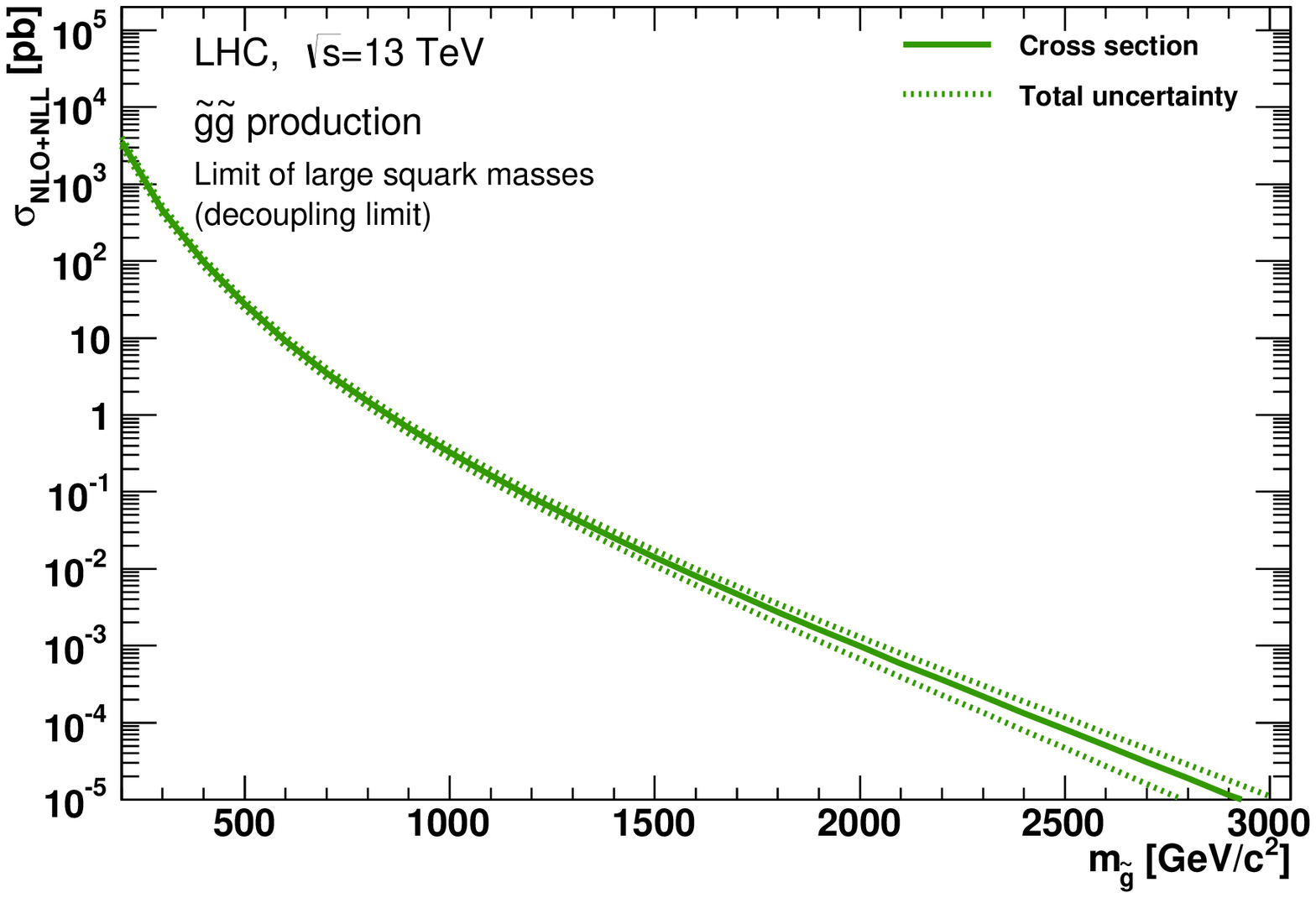}
       	\includegraphics[width=12cm]{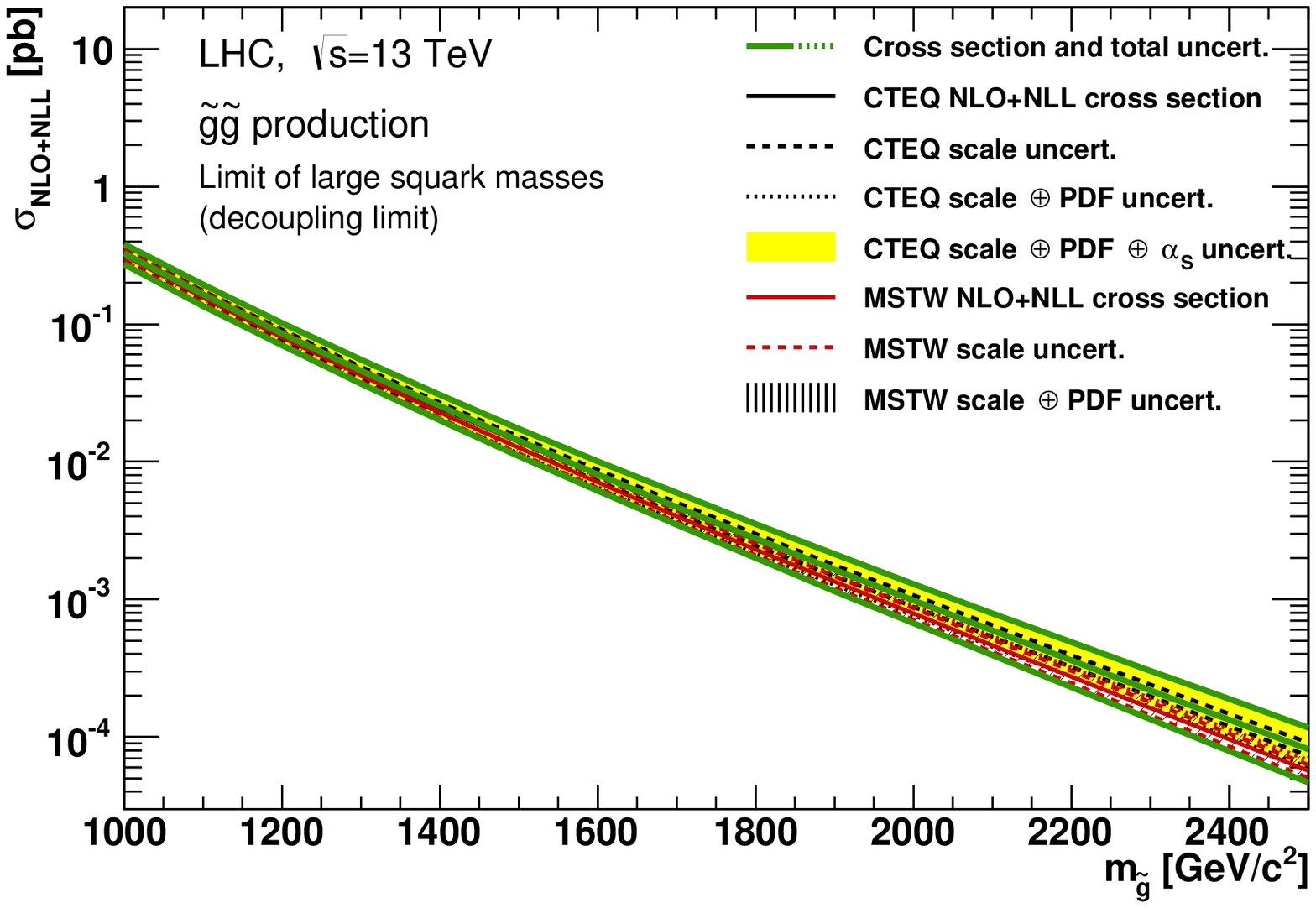}
  \end{center}
  \caption{\myxsectcaption{gluino pair}{squarks}{13}}
  \label{fig:xsect_gluino13}
\end{figure}

\begin{figure}[htbp]
  \begin{center}
       	\includegraphics[width=12cm]{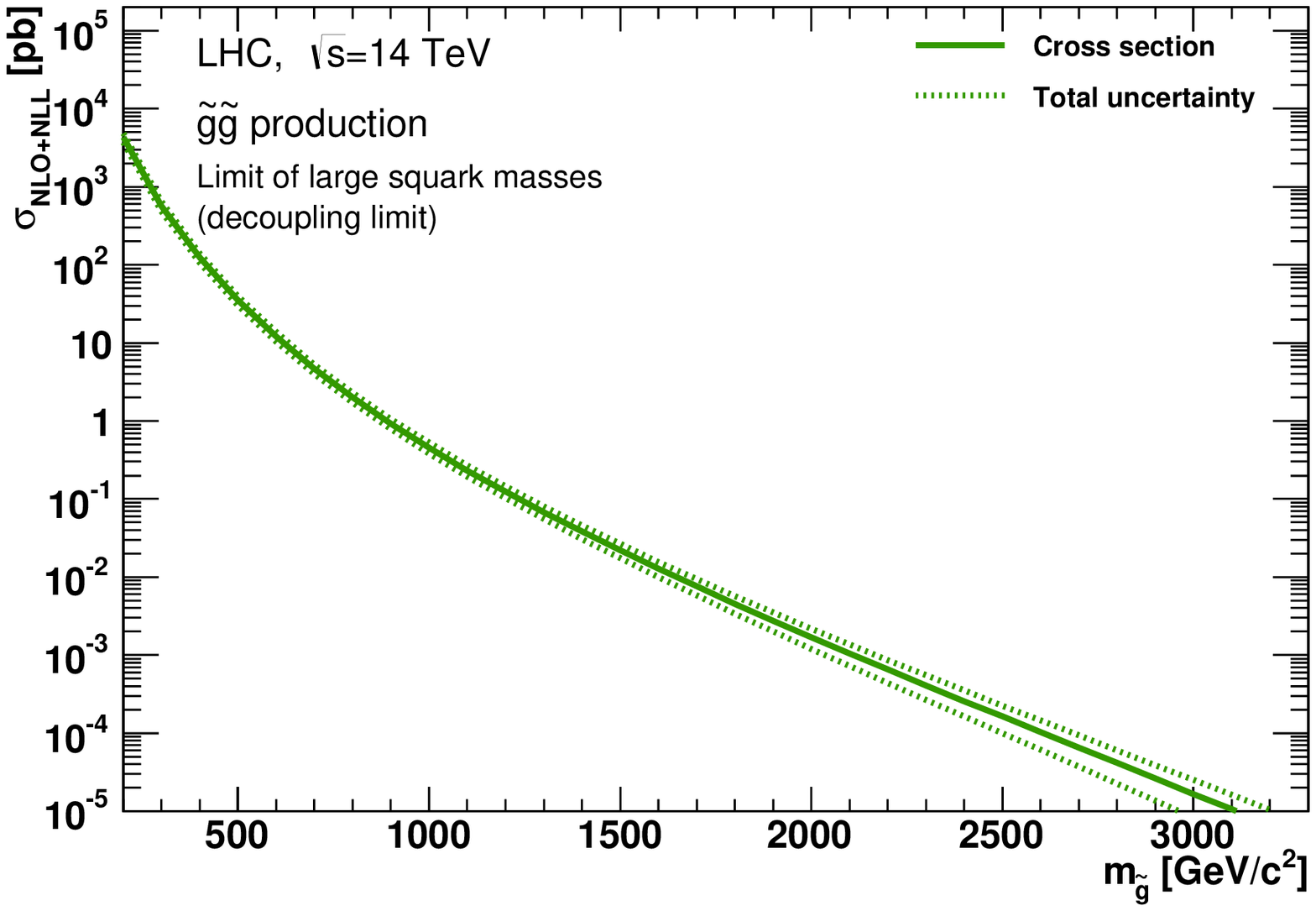}
       	\includegraphics[width=12cm]{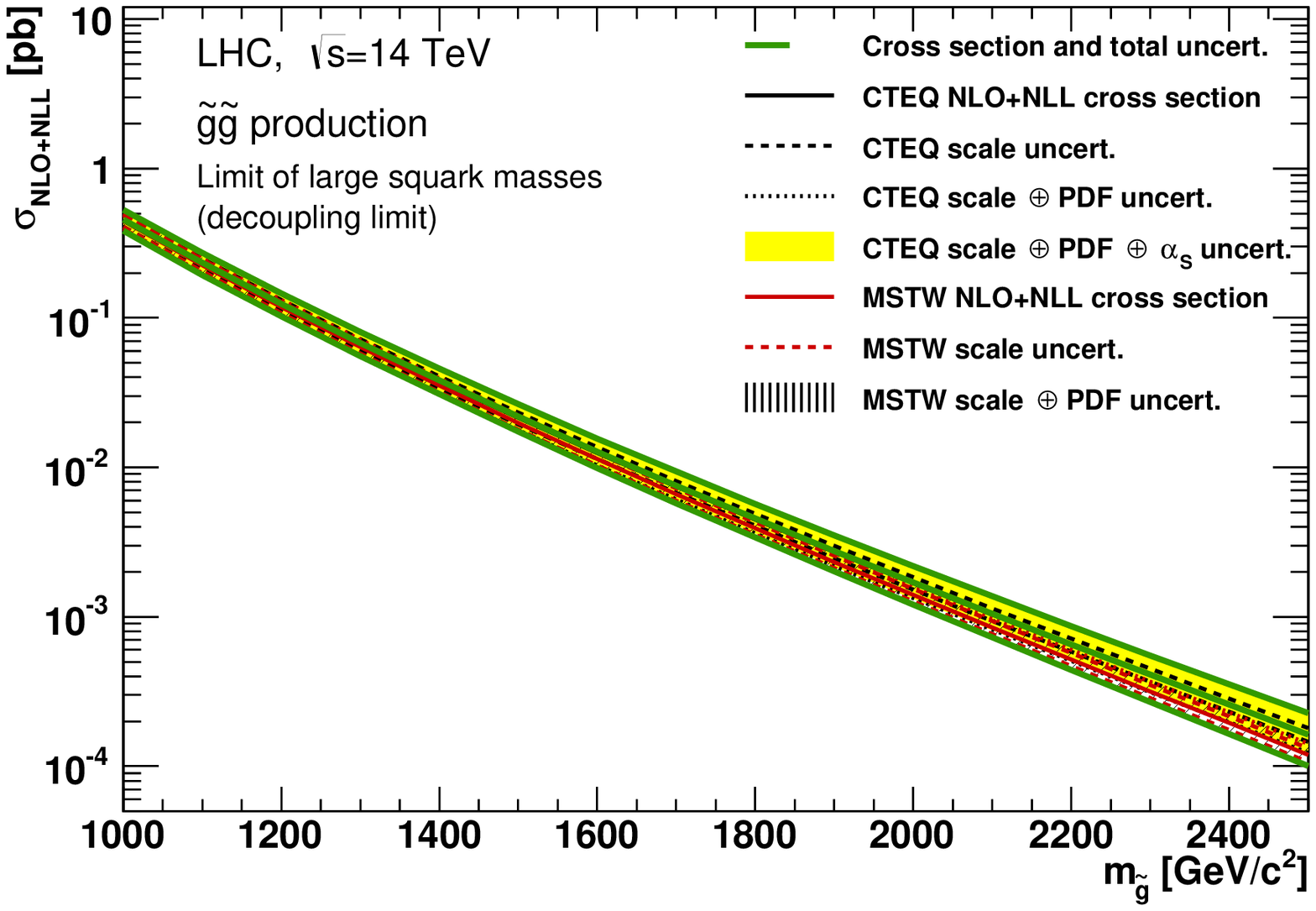}
  \end{center}
  \caption{\myxsectcaption{gluino pair}{squarks}{14}}
  \label{fig:xsect_gluino14}
\end{figure}

\begin{figure}[htbp]
  \begin{center}
       	\includegraphics[width=12cm]{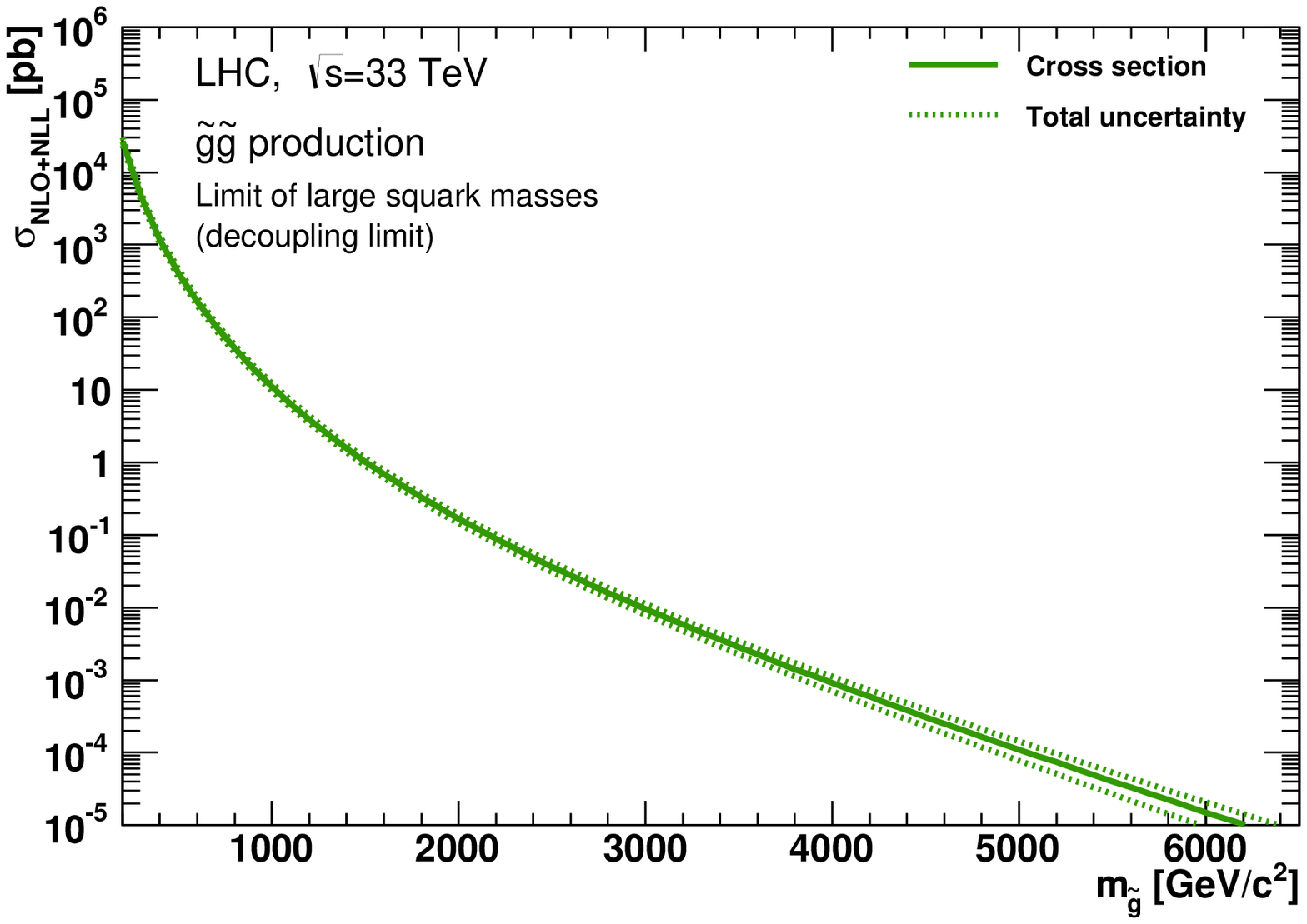}
       	\includegraphics[width=12cm]{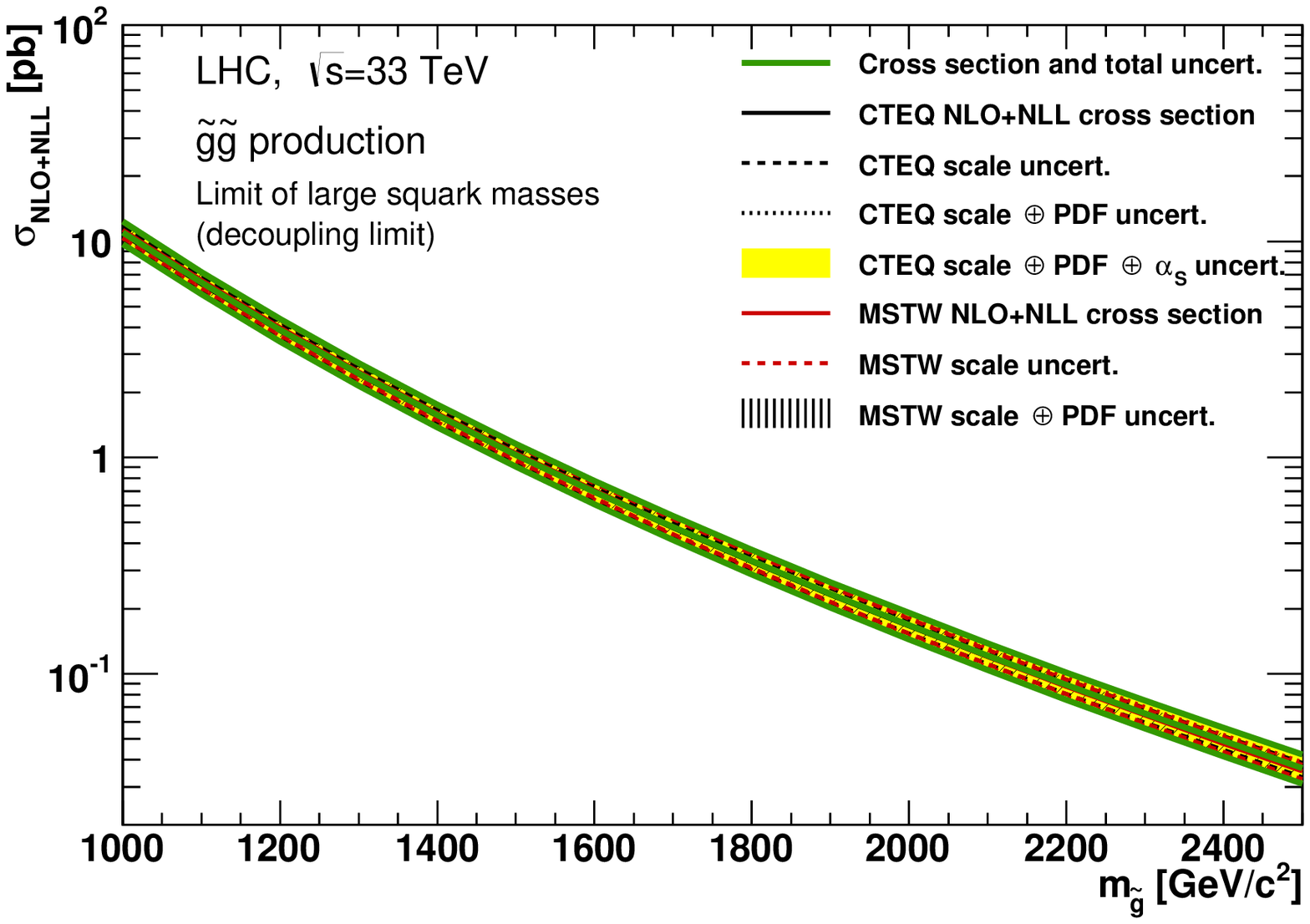}
  \end{center}
  \caption{\myxsectcaption{gluino pair}{squarks}{33}}
  \label{fig:xsect_gluino33}
\end{figure}

\begin{figure}[htbp]
  \begin{center}
       	\includegraphics[width=12cm]{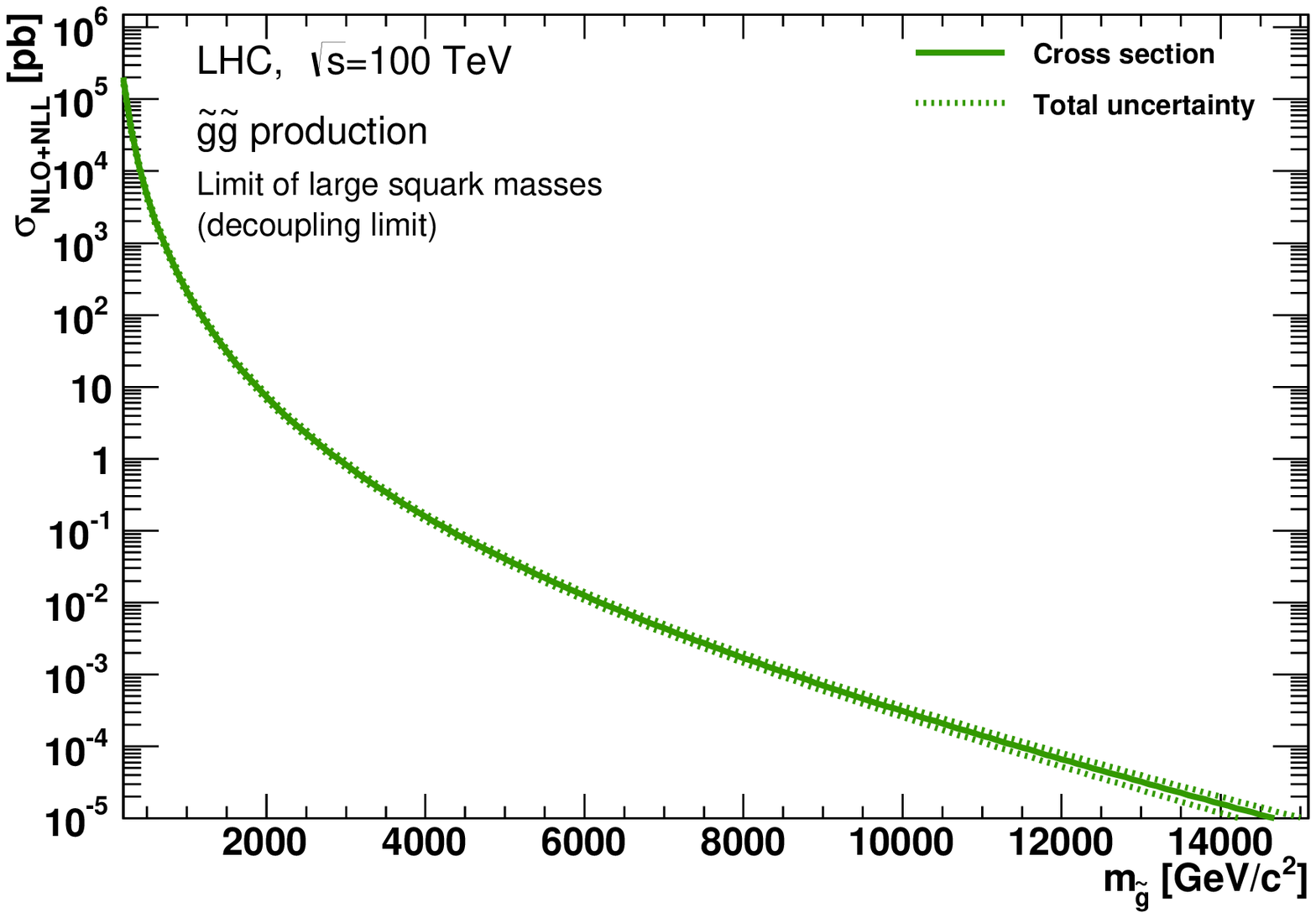}
       	\includegraphics[width=12cm]{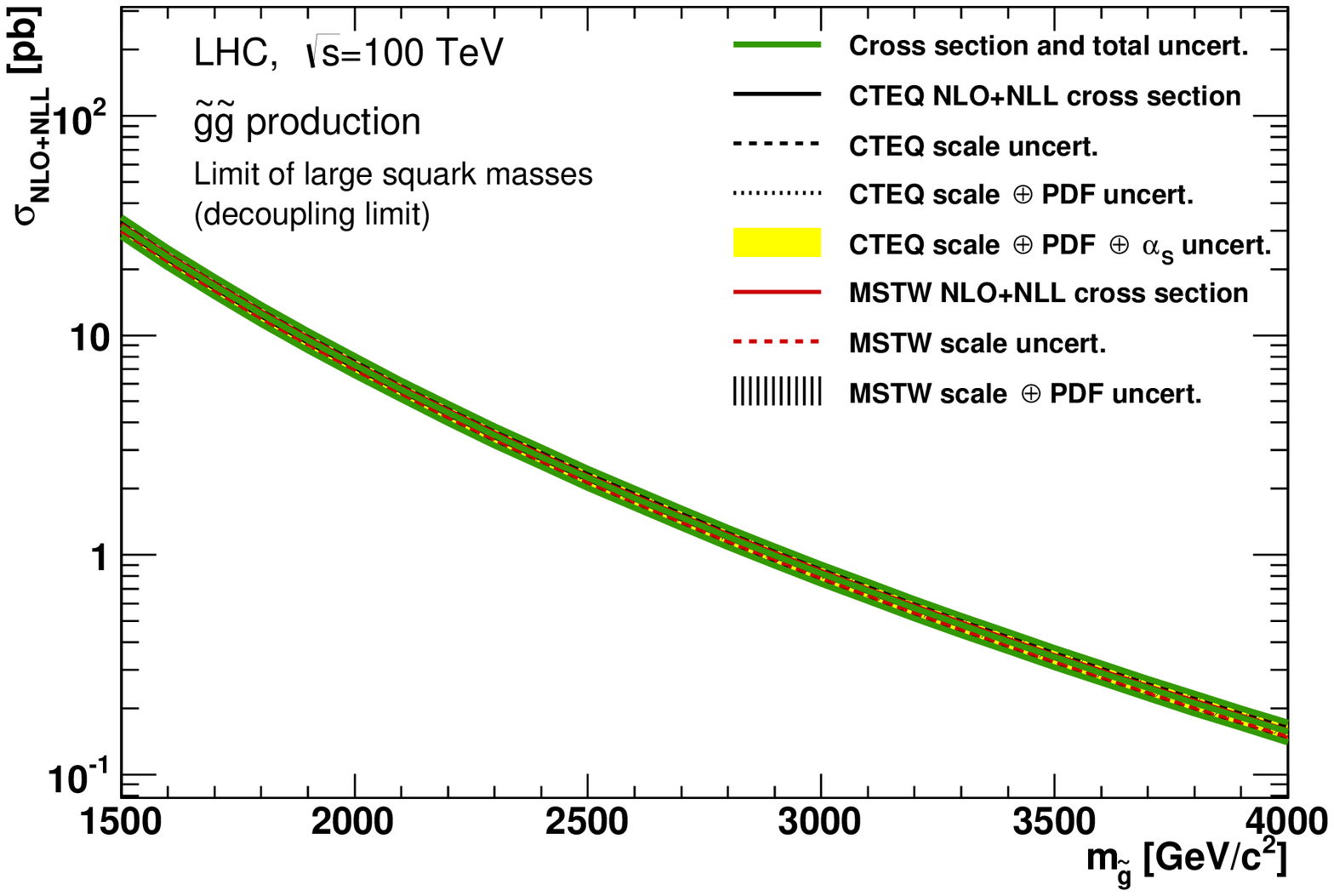}
  \end{center}
  \caption{\myxsectcaption{gluino pair}{squarks}{100}}
  \label{fig:xsect_gluino100}
\end{figure}

The gluino pair production cross section in a model where the squarks
are decoupled is shown in Figure~\ref{fig:xsect_gluino13}  for $\sqrt{s}=13$ TeV, Figure~\ref{fig:xsect_gluino14}  for $\sqrt{s}=14$ TeV, in Figure~\ref{fig:xsect_gluino33}  for $\sqrt{s}=33$ TeV and in Figure~\ref{fig:xsect_gluino100}  for $\sqrt{s}=100$ TeV. The gluino
mass spans the range from 0.2 TeV to 3 TeV in the upper plot of Figure~\ref{fig:xsect_gluino13}. The close up of the mass range from 1 to 2.5 TeV for $\sqrt{s}=13$ TeV is provided in the lower plot of Figure~\ref{fig:xsect_gluino13}. Results for $\sqrt{s}=14$ TeV in the mass ranges $0.2-3.3$ TeV and  $1-2.5$ TeV are shown in  Figure~\ref{fig:xsect_gluino14} and results for $\sqrt{s}=33$ TeV in the mass ranges $0.2-6.5$ TeV and $1-2.5$ TeV are presented in Figure~\ref{fig:xsect_gluino33}, respectively.  Similarly,  Figure~\ref{fig:xsect_gluino100} shows results for $\sqrt{s}=100$ TeV in the mass ranges $0.2-15$ TeV and $1.5-4$ TeV.  In the lower figures, the black (red) line corresponds to
the NLO+NLL nominal cross section and renormalisation and
factorisation scale uncertainties obtained using the \cteq\ (\mstw)
PDF set. The solid yellow (dashed black) band corresponds to the total
uncertainty of the cross section using \cteq\ (\mstw), as derived from
Eq.~(\ref{eq:variations}). Finally, the green lines in the upper and lower plots delimit the
envelope and the central value. They correspond to the central nominal
value together with the total uncertainties.

\subsection{Squark-antisquark production}

\begin{figure}[htbp]
  \begin{center}
       	\includegraphics[width=12cm]{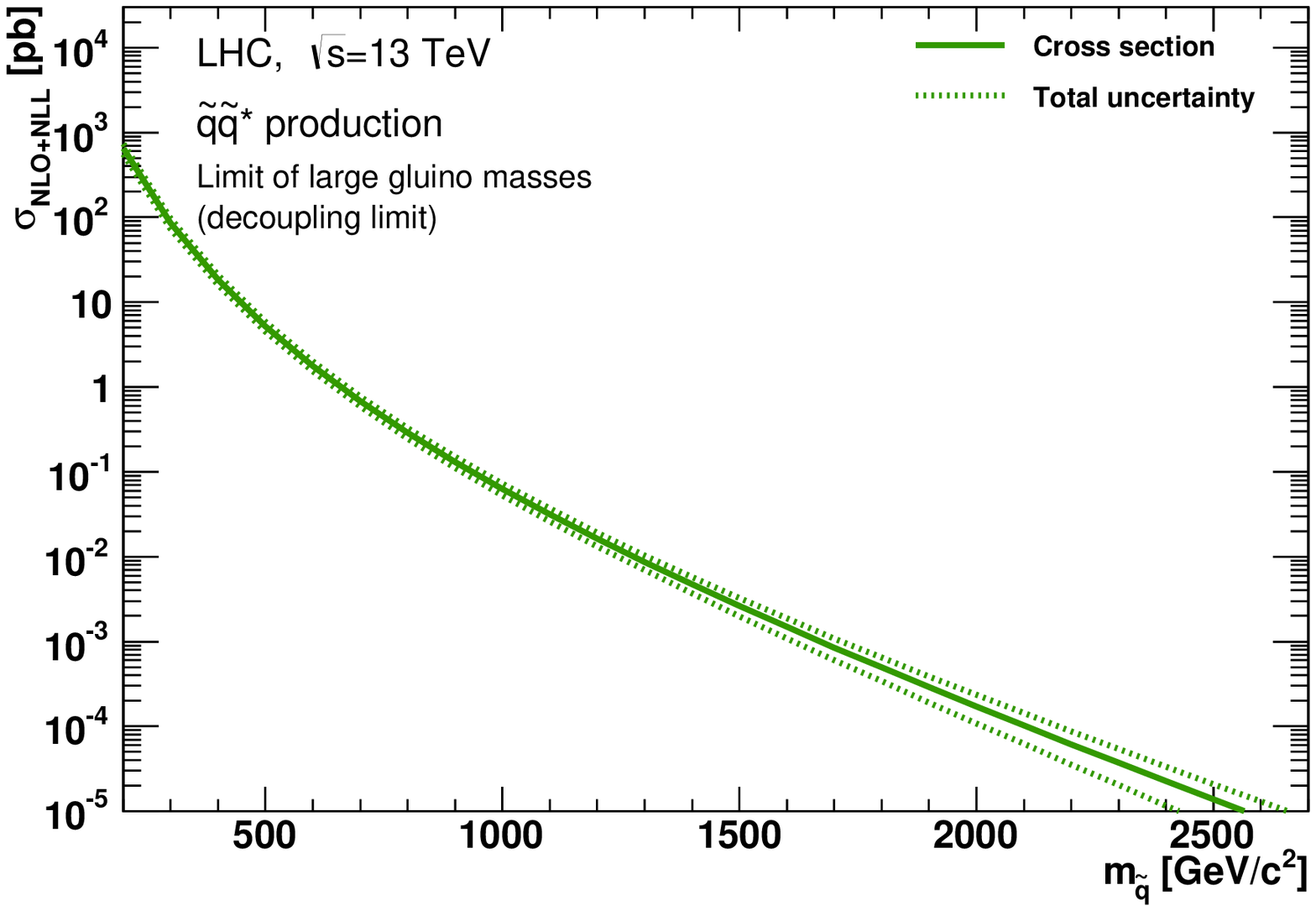}
       	\includegraphics[width=12cm]{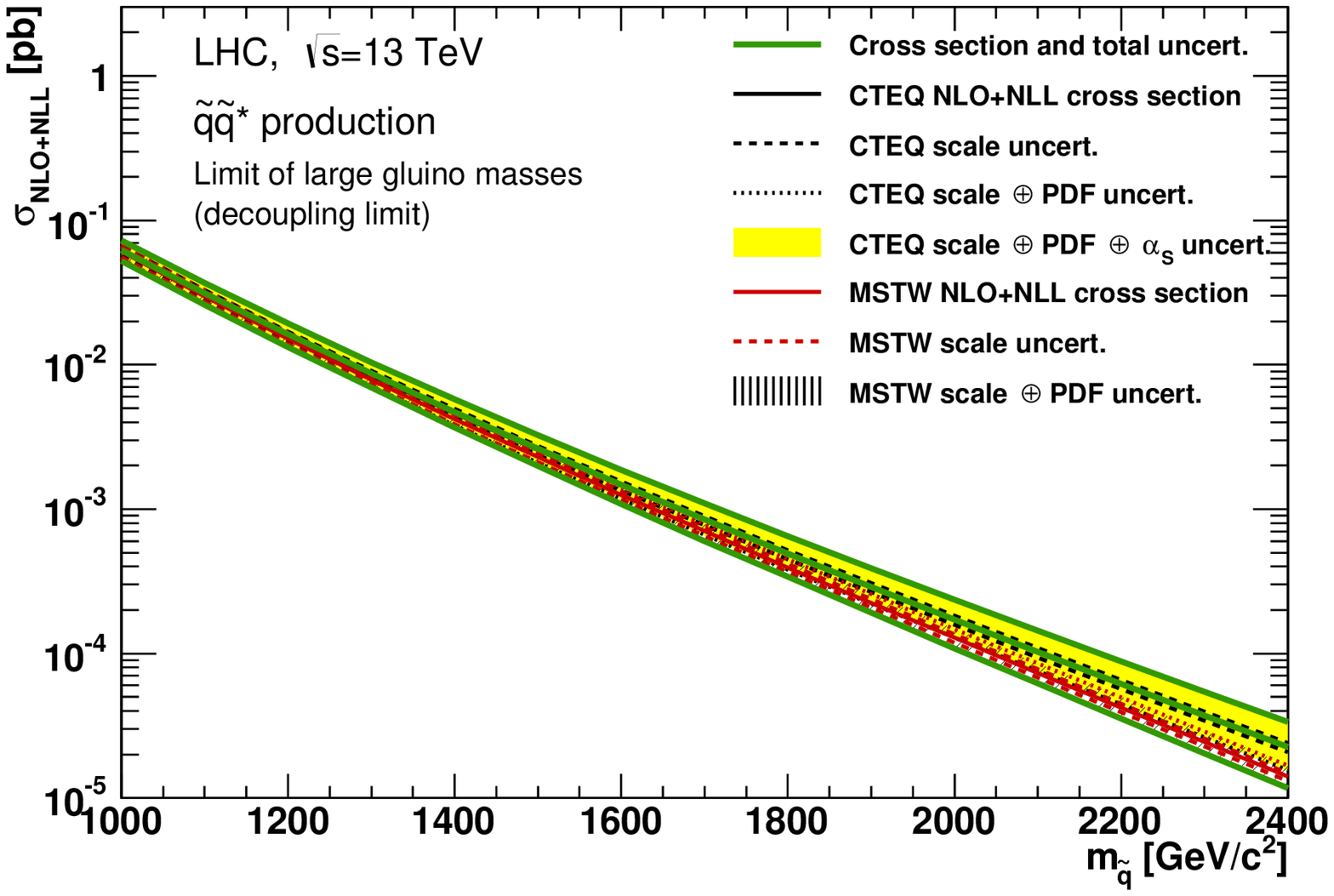}
  \end{center}
  \caption{\myxsectcaption{squark-antisquark}{gluinos}{13}}
  \label{fig:xsect_squark13}
\end{figure}

\begin{figure}[htbp]
  \begin{center}
       	\includegraphics[width=12cm]{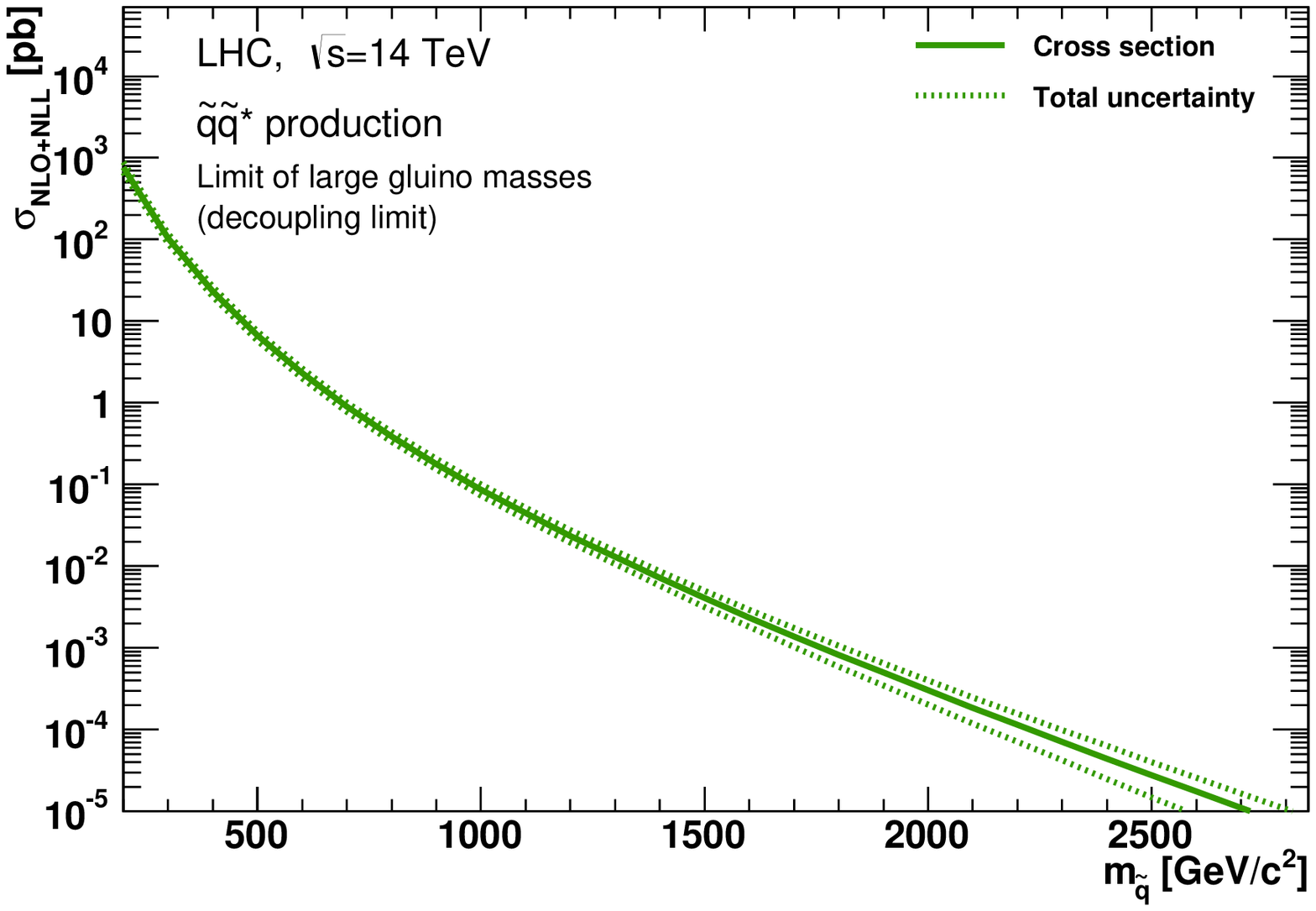}
       	\includegraphics[width=12cm]{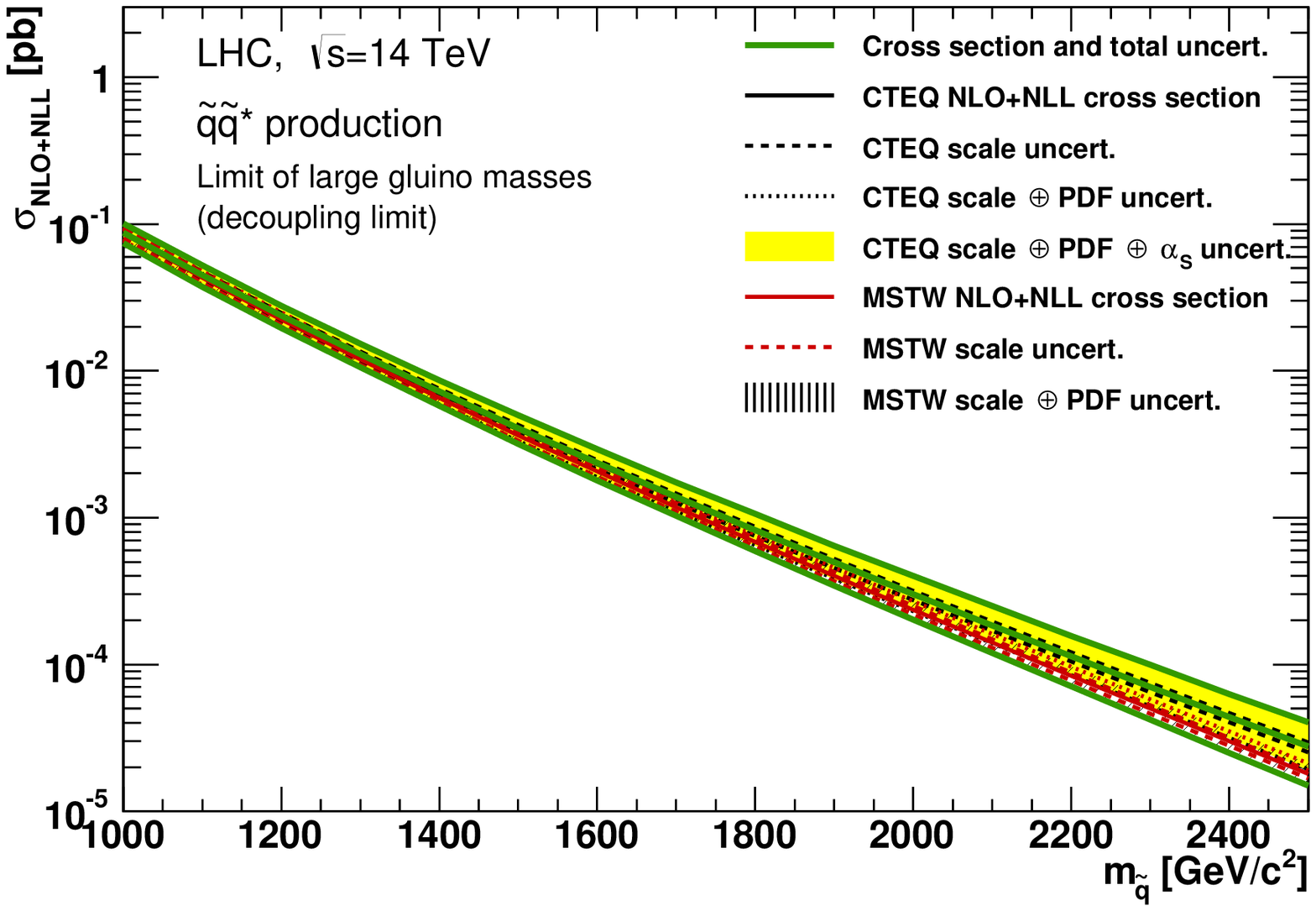}
  \end{center}
  \caption{\myxsectcaption{squark-antisquark}{gluinos}{14}}
  \label{fig:xsect_squark14}
\end{figure}

\begin{figure}[htbp]
  \begin{center}
       	\includegraphics[width=12cm]{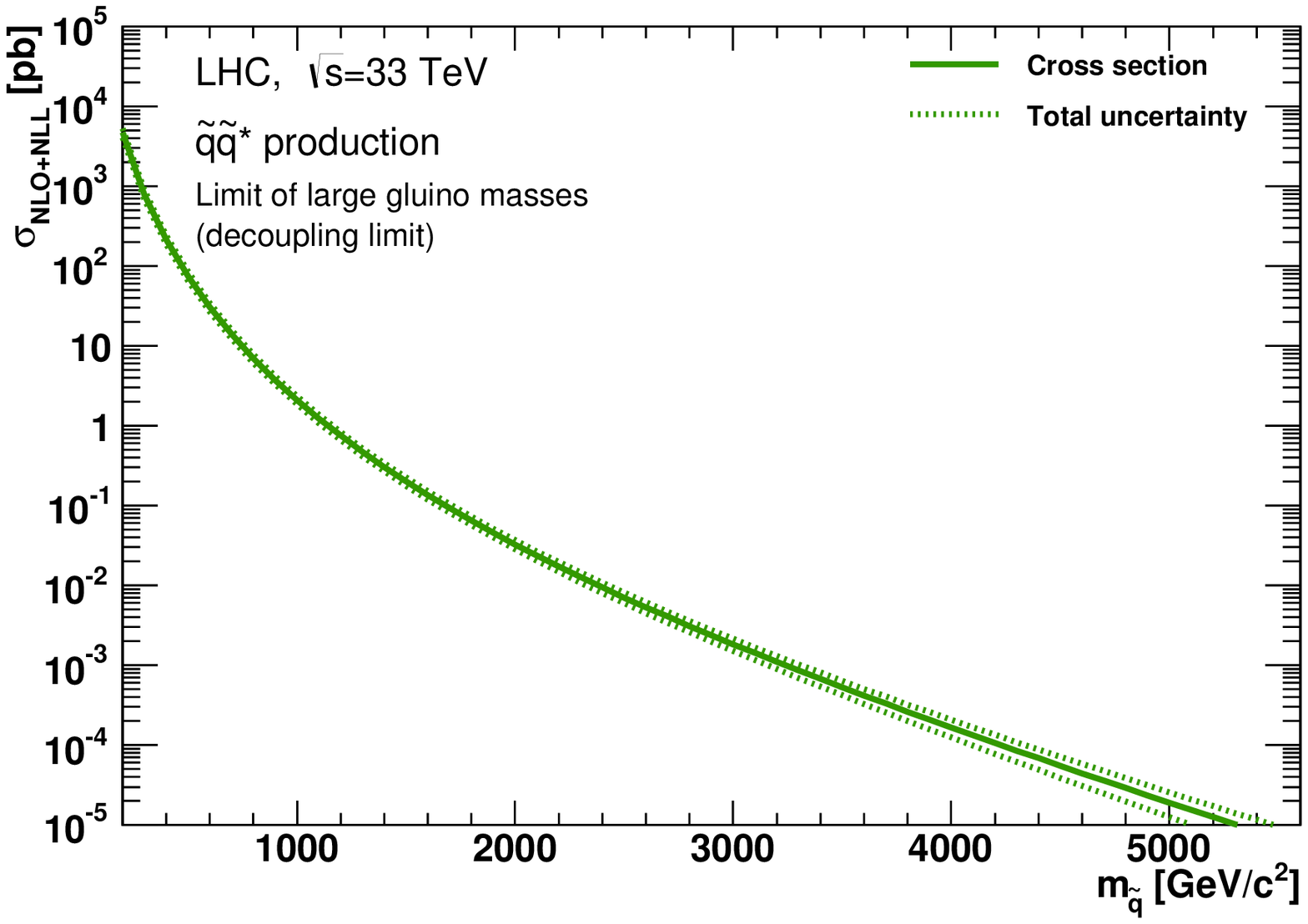}
       	\includegraphics[width=12cm]{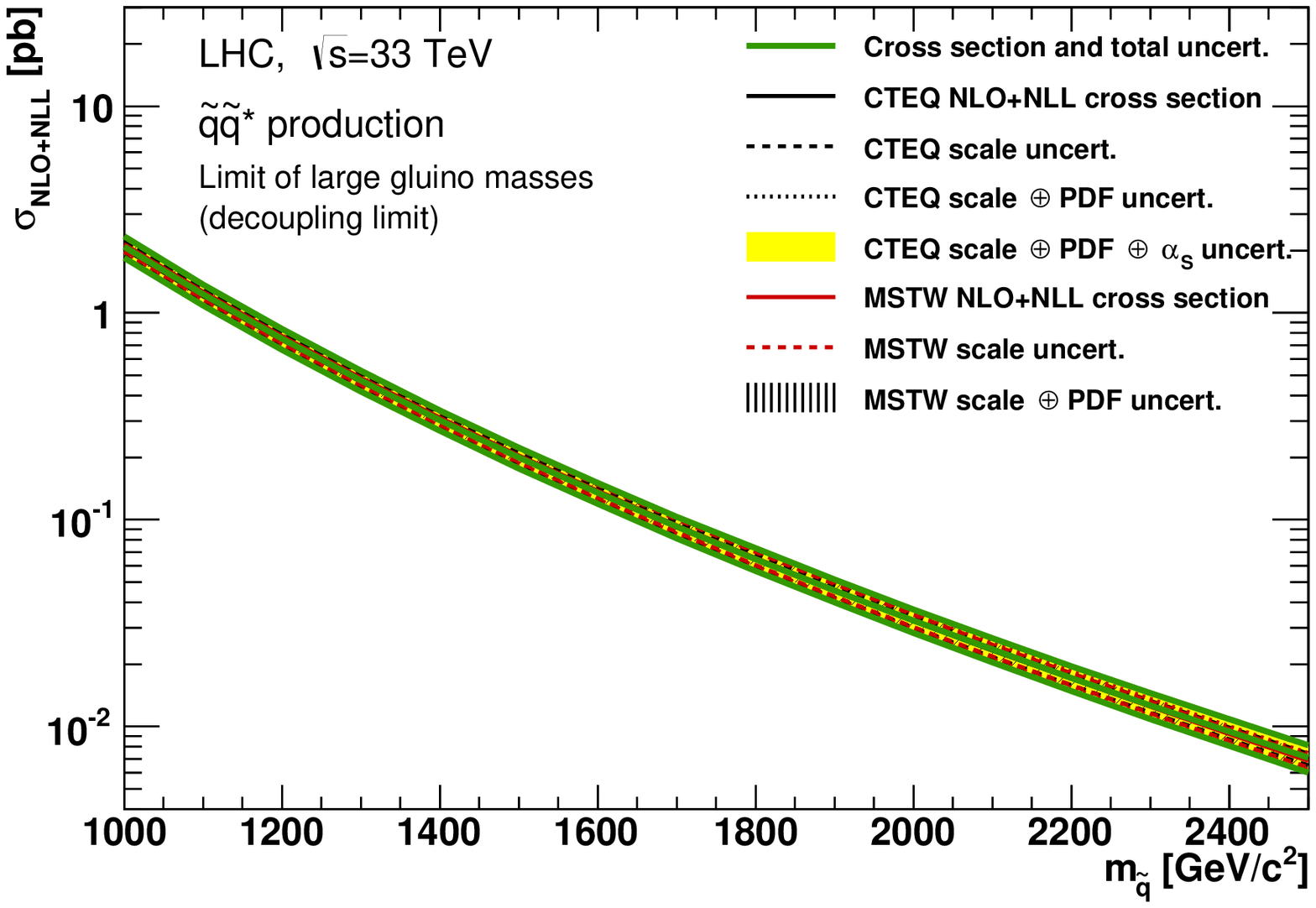}
  \end{center}
  \caption{\myxsectcaption{squark-antisquark}{gluinos}{33}}
  \label{fig:xsect_squark33}
\end{figure}

\begin{figure}[htbp]
  \begin{center}
       	\includegraphics[width=12cm]{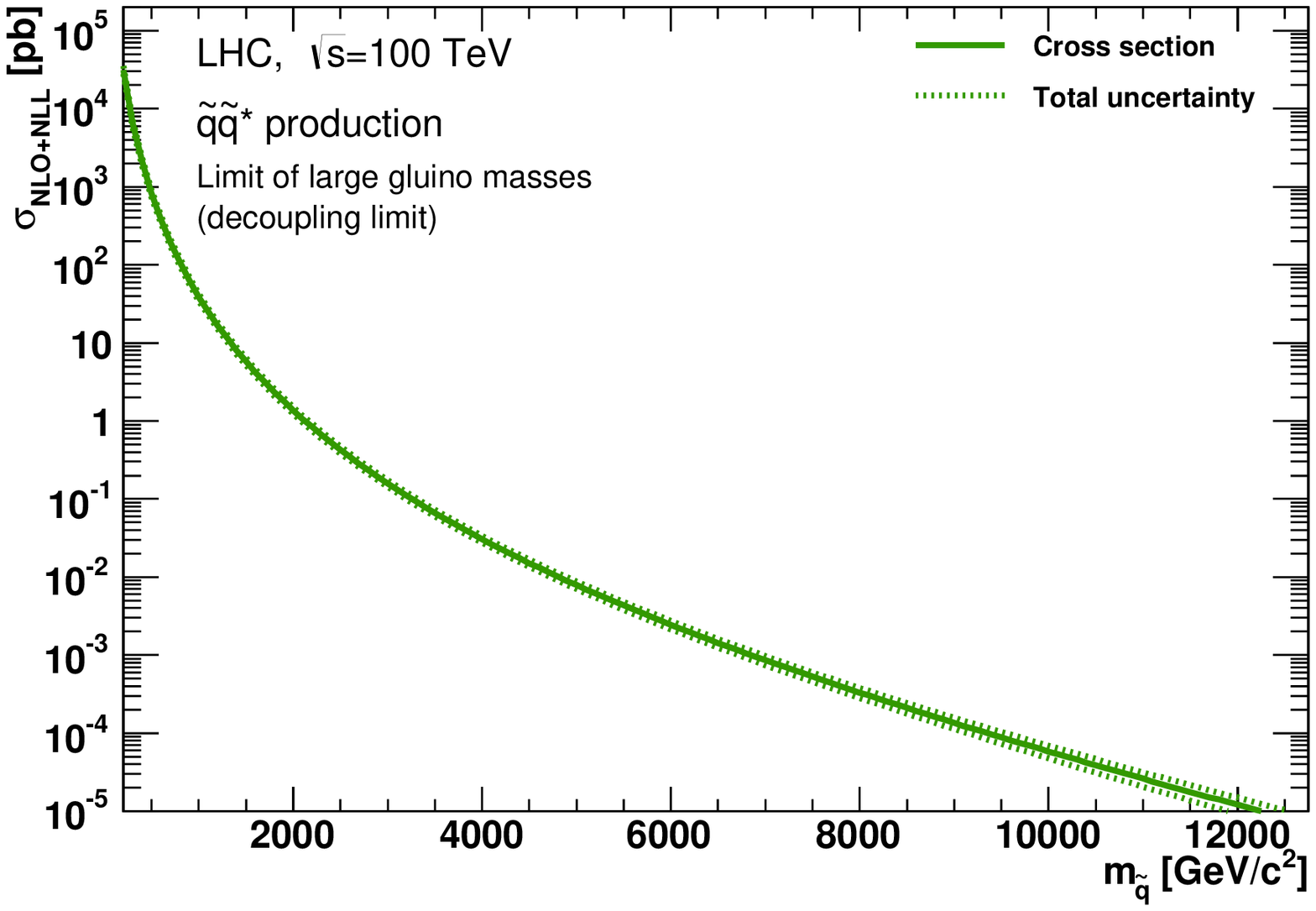}
       	\includegraphics[width=12cm]{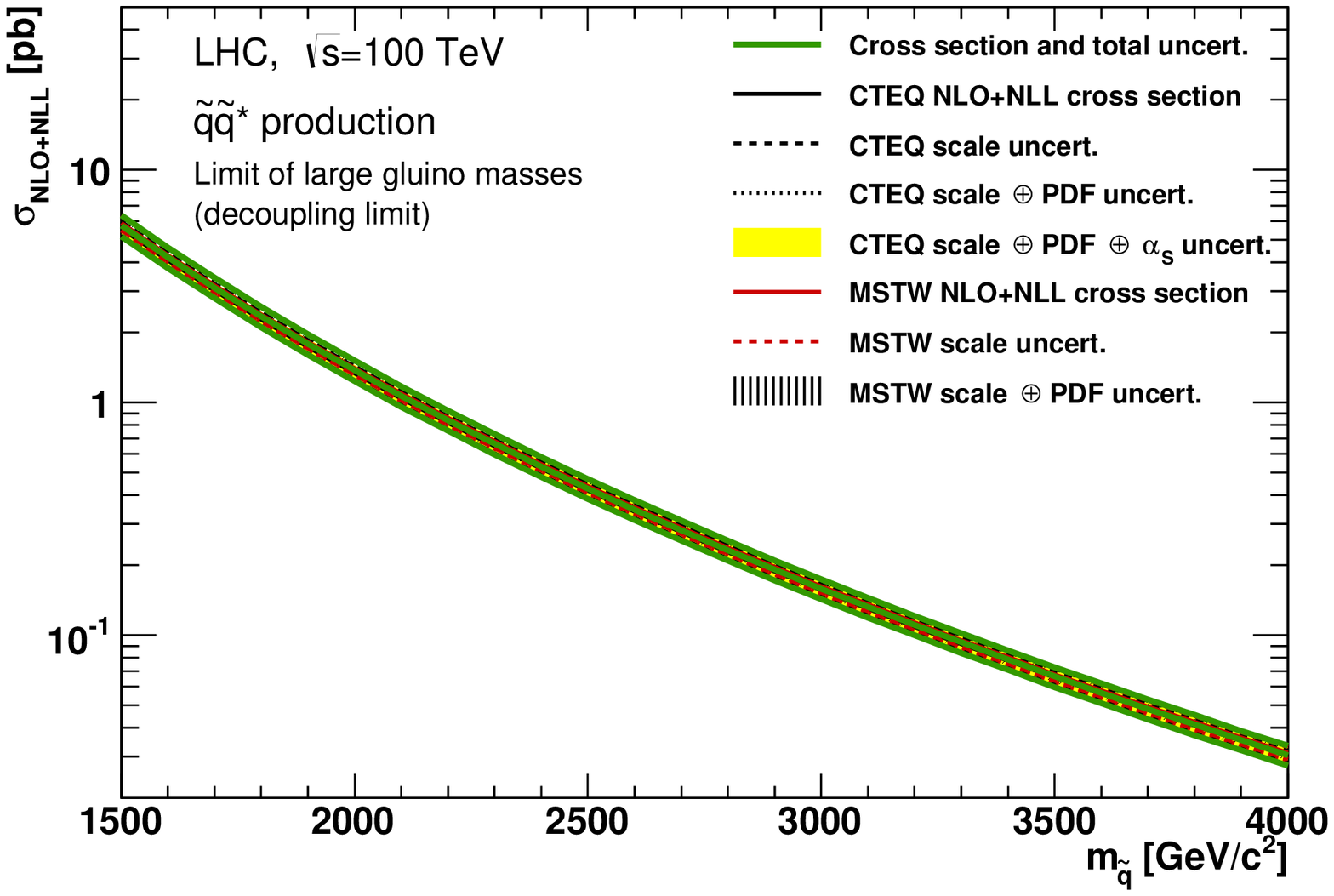}
  \end{center}
  \caption{\myxsectcaption{squark-antisquark}{gluinos}{100}}
  \label{fig:xsect_squark100}
\end{figure}

In order to show the evolution of the squark-antisquark production
cross section as a function of the squark mass, a scenario has been
chosen in which the gluino is decoupled. The results are shown in
Figures~\ref{fig:xsect_squark13},~\ref{fig:xsect_squark14},~\ref{fig:xsect_squark33} and in
~\ref{fig:xsect_squark100}, using the same convention for the display 
of the various contributions as in the gluino pair production case.  

\subsection{Direct stop and sbottom pair production}

\begin{figure}[htbp]
  \begin{center}
       	\includegraphics[width=12cm]{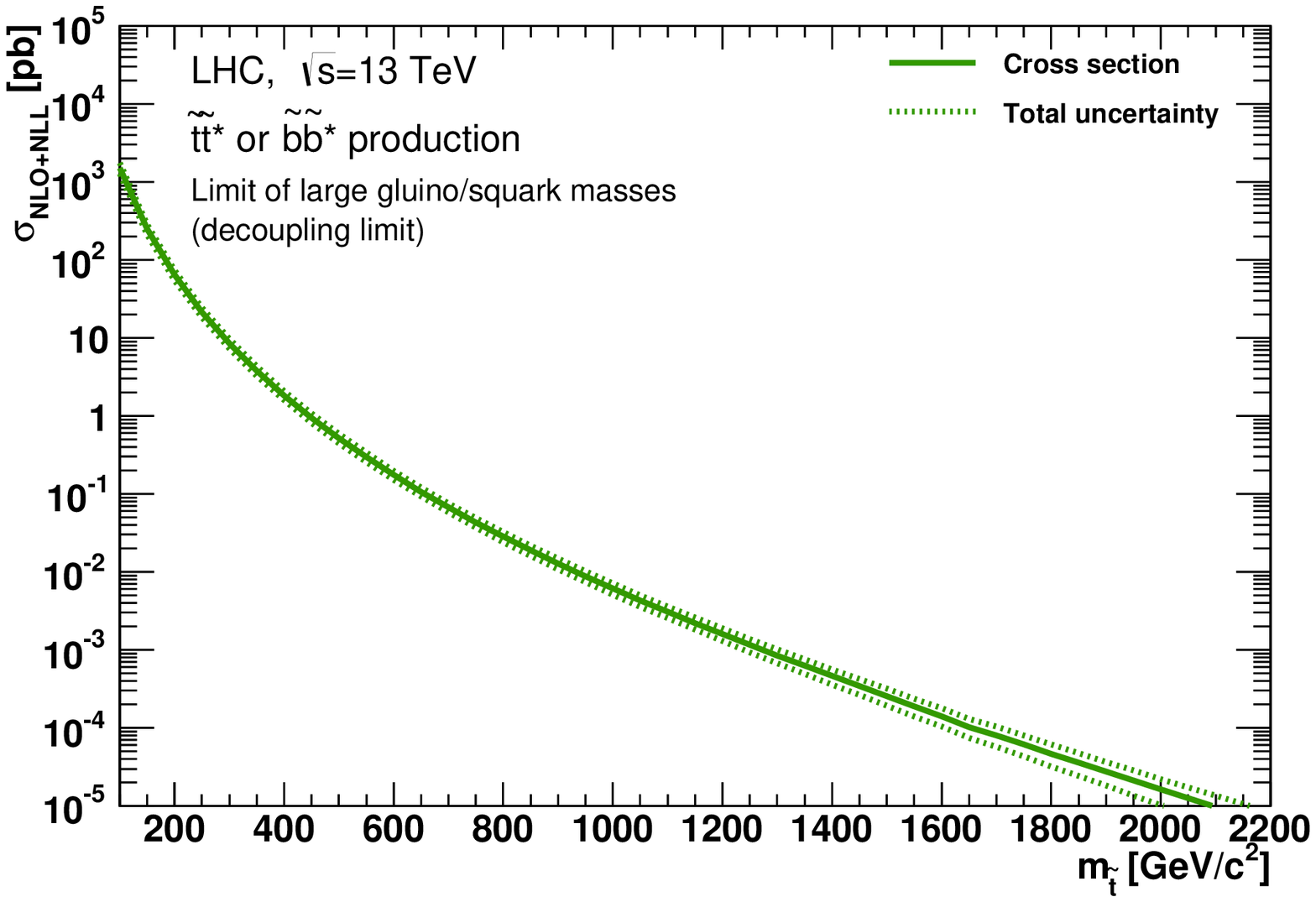}
       	\includegraphics[width=12cm]{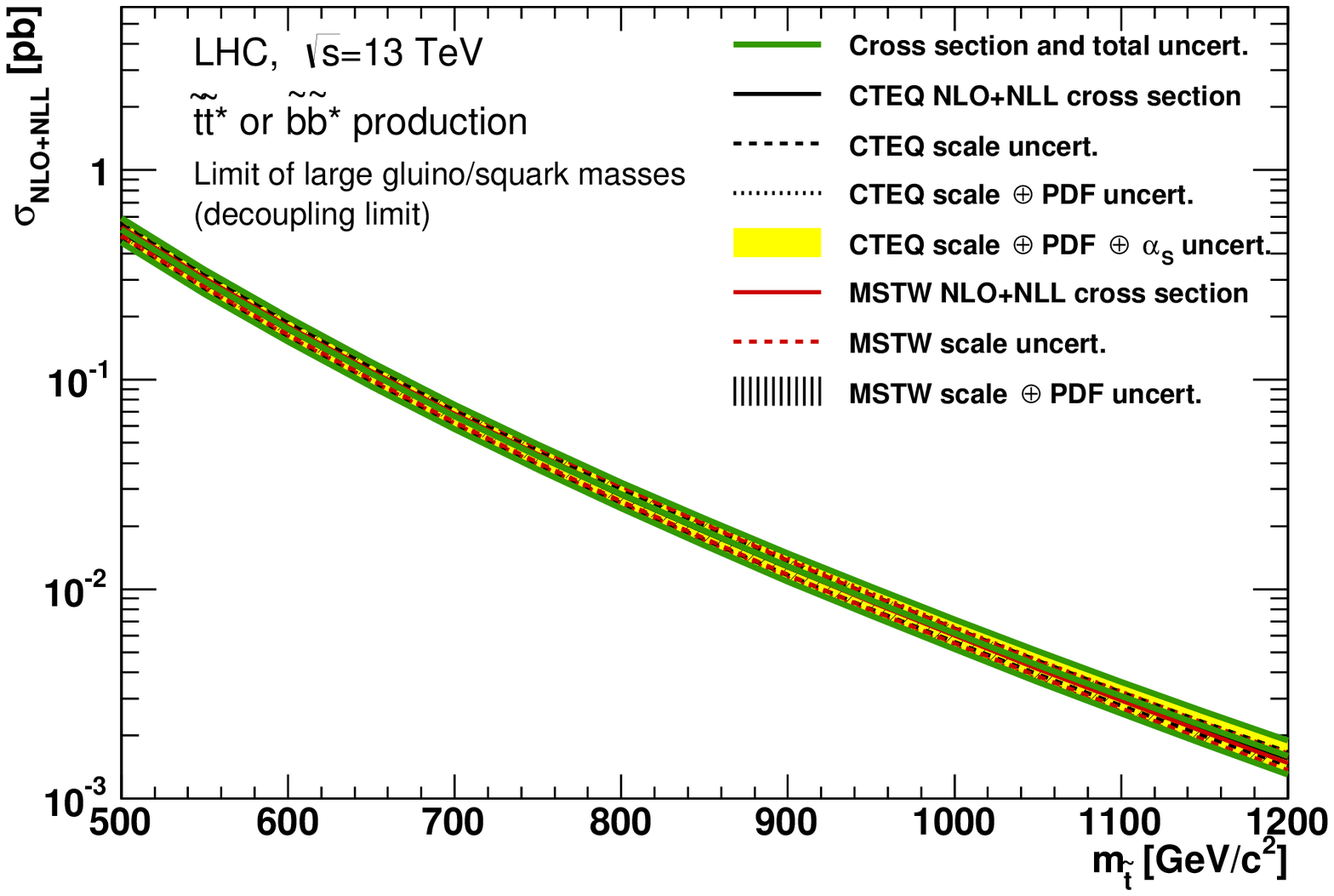}
  \end{center}
  \caption{NLO+NLL stop-antistop production cross section as a function of mass at $\sqrt{s}=13$ TeV in the wider (upper plot) and narrower (lower plot) mass range. The different 
   styled black (red) lines correspond to the cross section and scale uncertainties predicted
  using the \cteq\ (\mstw) PDF set. The yellow (dashed black) band
  corresponds to the total \cteq\ (\mstw) uncertainty, as described in
  the text. The green lines show the final cross section and its total
  uncertainty.}
  \label{fig:xsect_stop13}
\end{figure}

\begin{figure}[htbp]
  \begin{center}
       	\includegraphics[width=12cm]{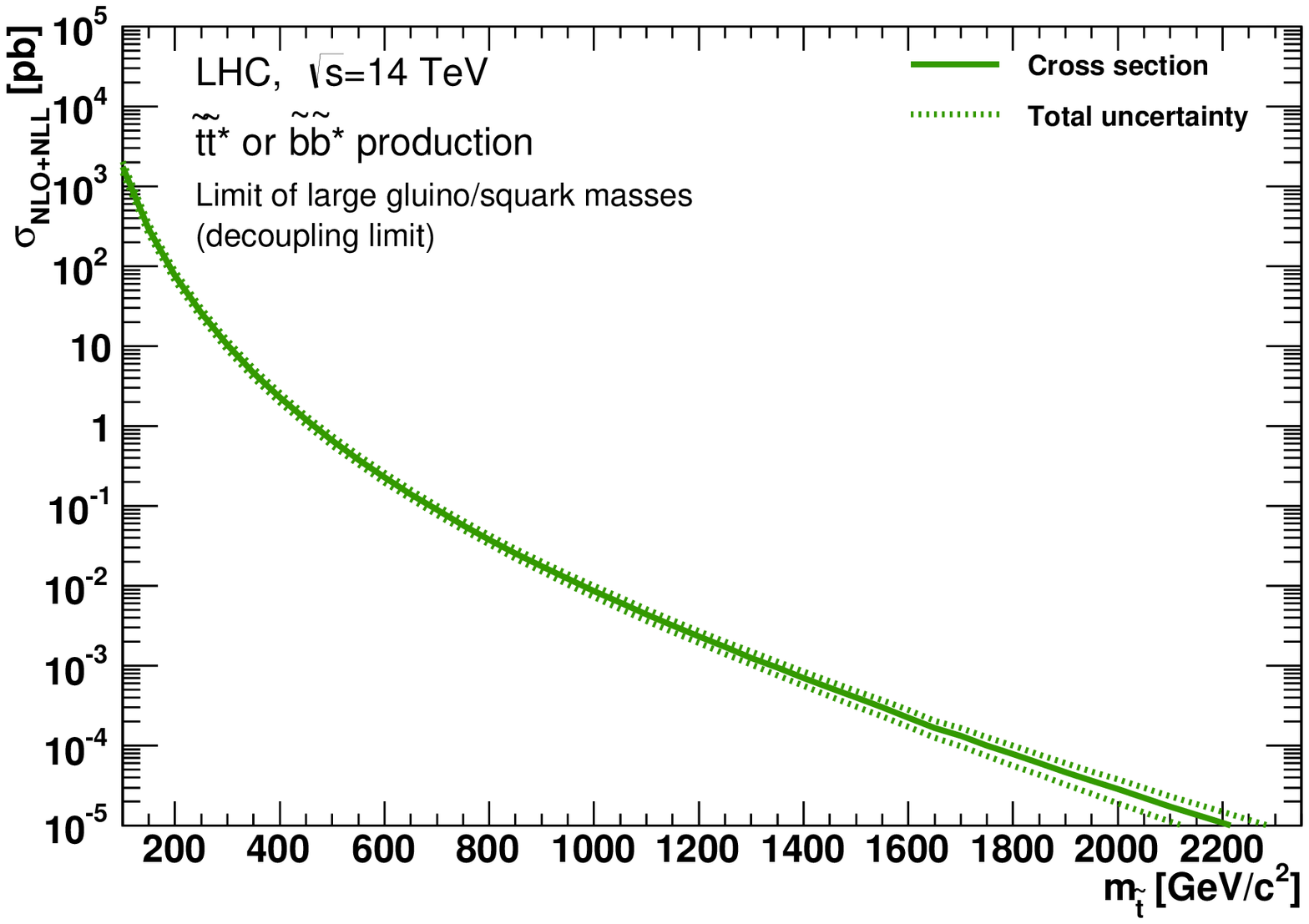}
       	\includegraphics[width=12cm]{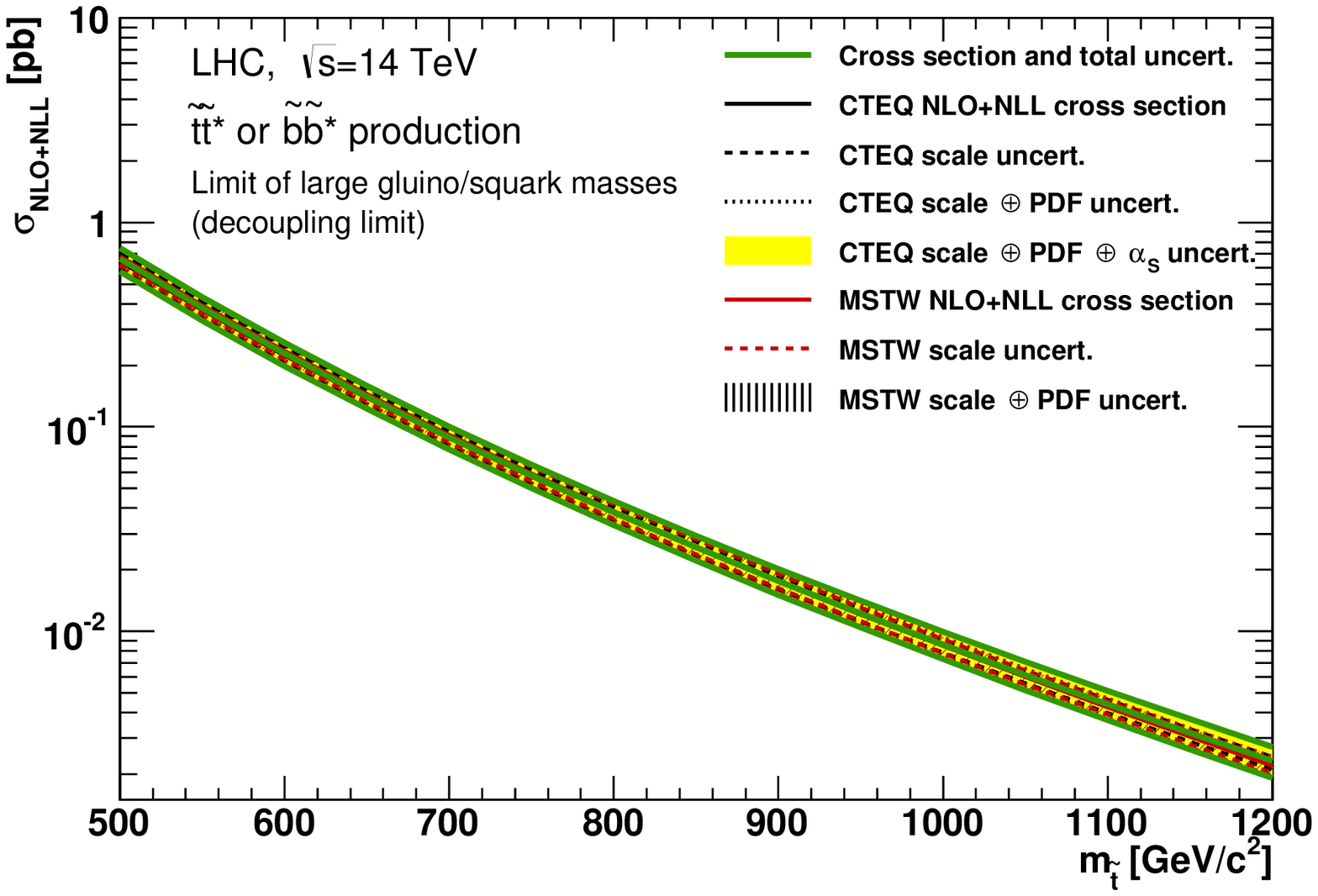}
  \end{center}
  \caption{NLO+NLL stop-antistop production cross section as a function of mass at $\sqrt{s}=14$ TeV in the wider (upper plot) and narrower (lower plot) mass range. The different 
   styled black (red) lines correspond to the cross section and scale uncertainties predicted
  using the \cteq\ (\mstw) PDF set. The yellow (dashed black) band
  corresponds to the total \cteq\ (\mstw) uncertainty, as described in
  the text. The green lines show the final cross section and its total
  uncertainty.}
  \label{fig:xsect_stop14}
\end{figure}

\begin{figure}[htbp]
  \begin{center}
       	\includegraphics[width=12cm]{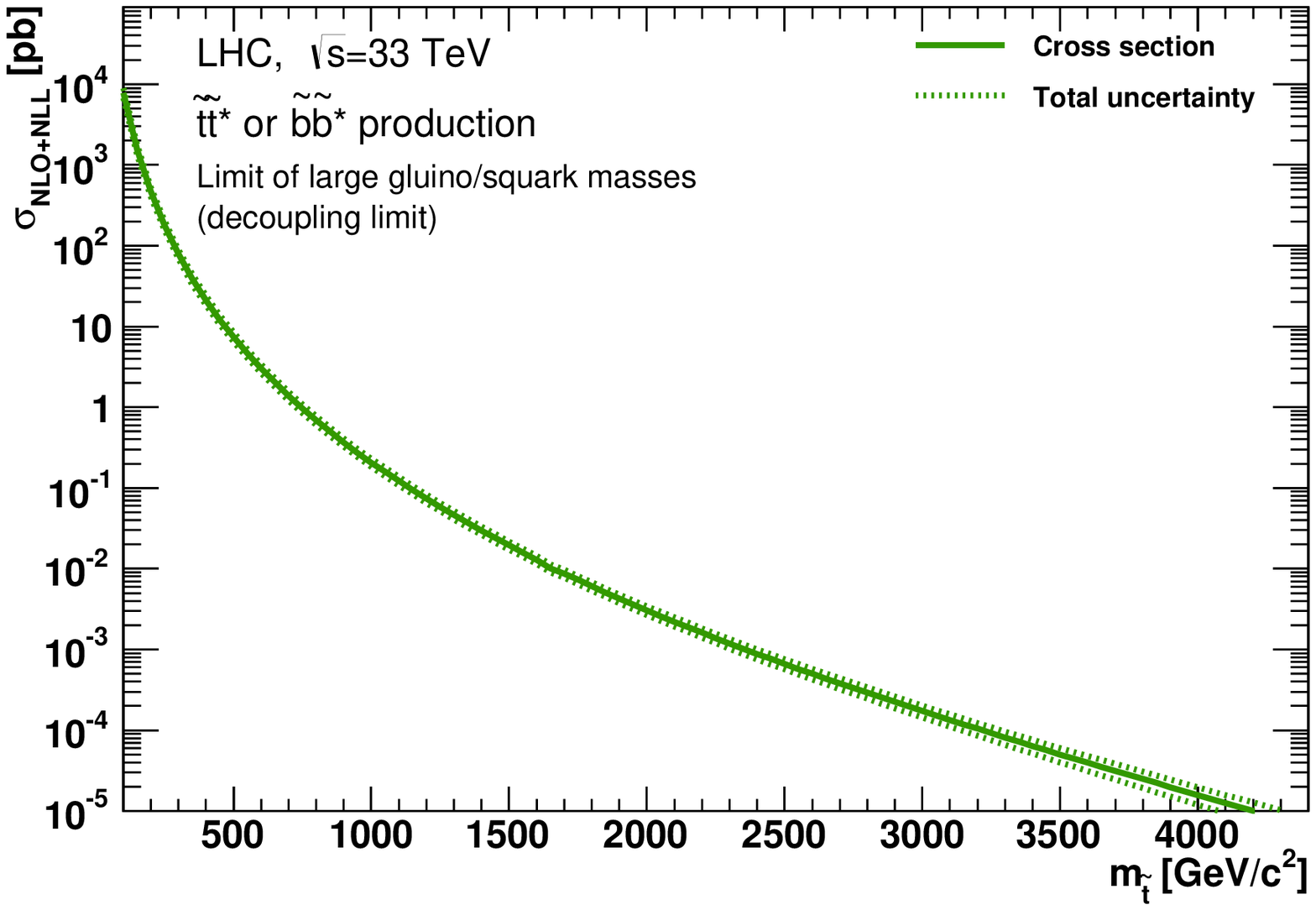}
       	\includegraphics[width=12cm]{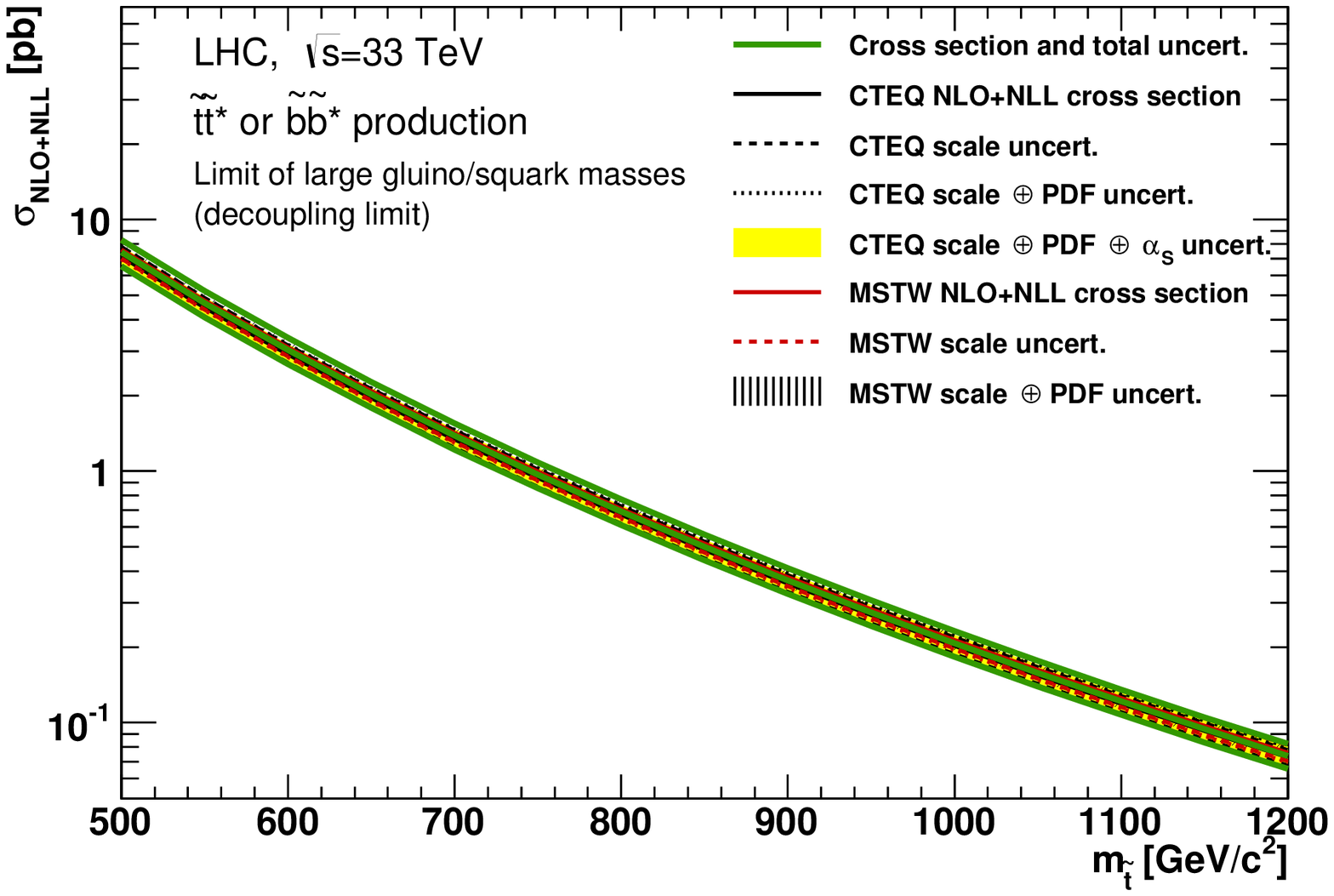}
  \end{center}
  \caption{NLO+NLL stop-antistop production cross section as a function of mass at $\sqrt{s}=33$ TeV in the wider (upper plot) and narrower (lower plot) mass range. The different 
   styled black (red) lines correspond to the cross section and scale uncertainties predicted
  using the \cteq\ (\mstw) PDF set. The yellow (dashed black) band
  corresponds to the total \cteq\ (\mstw) uncertainty, as described in
  the text. The green lines show the final cross section and its total
  uncertainty.}
  \label{fig:xsect_stop33}
\end{figure}

\begin{figure}[htbp]
  \begin{center}
       	\includegraphics[width=12cm]{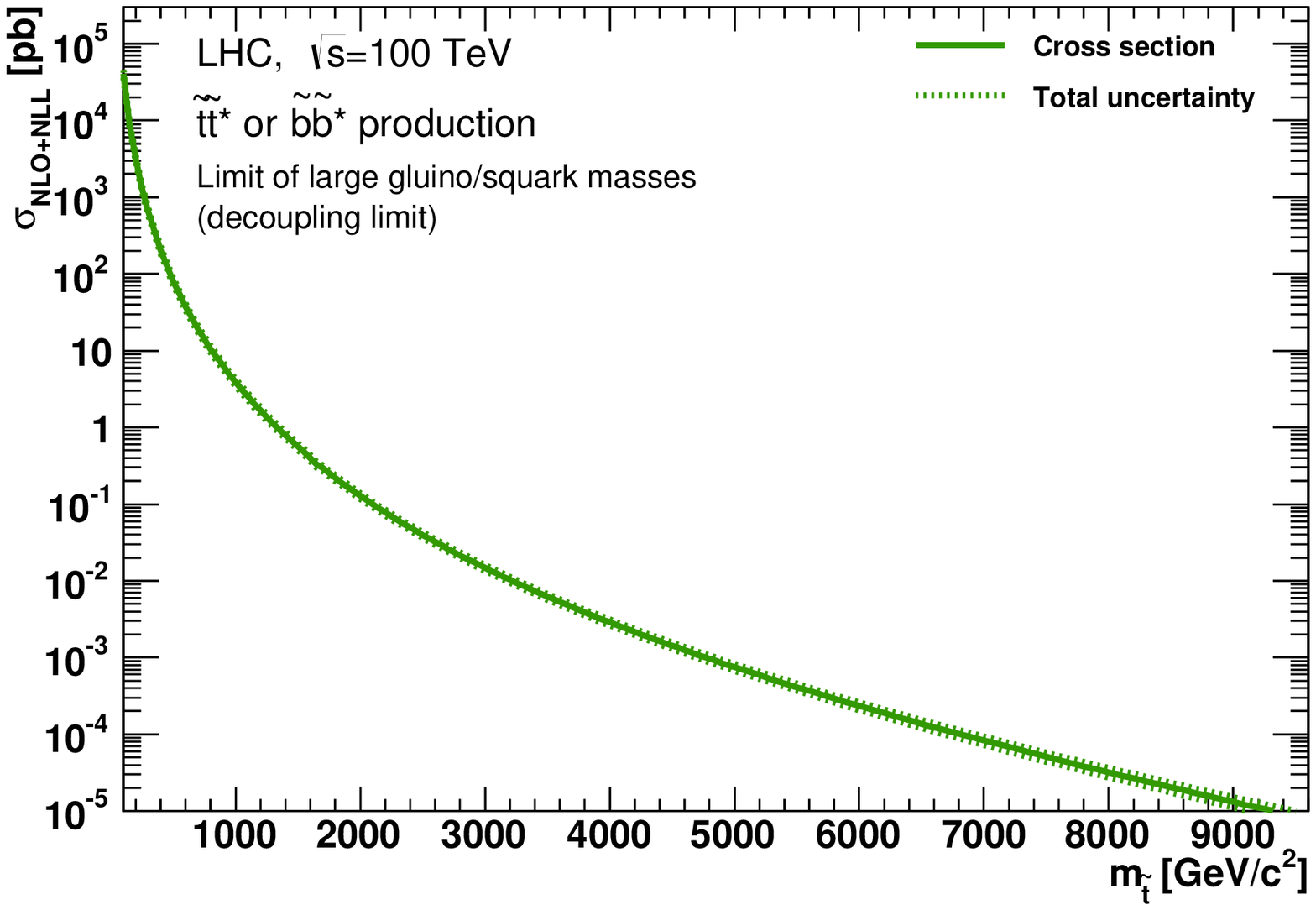}
       	\includegraphics[width=12cm]{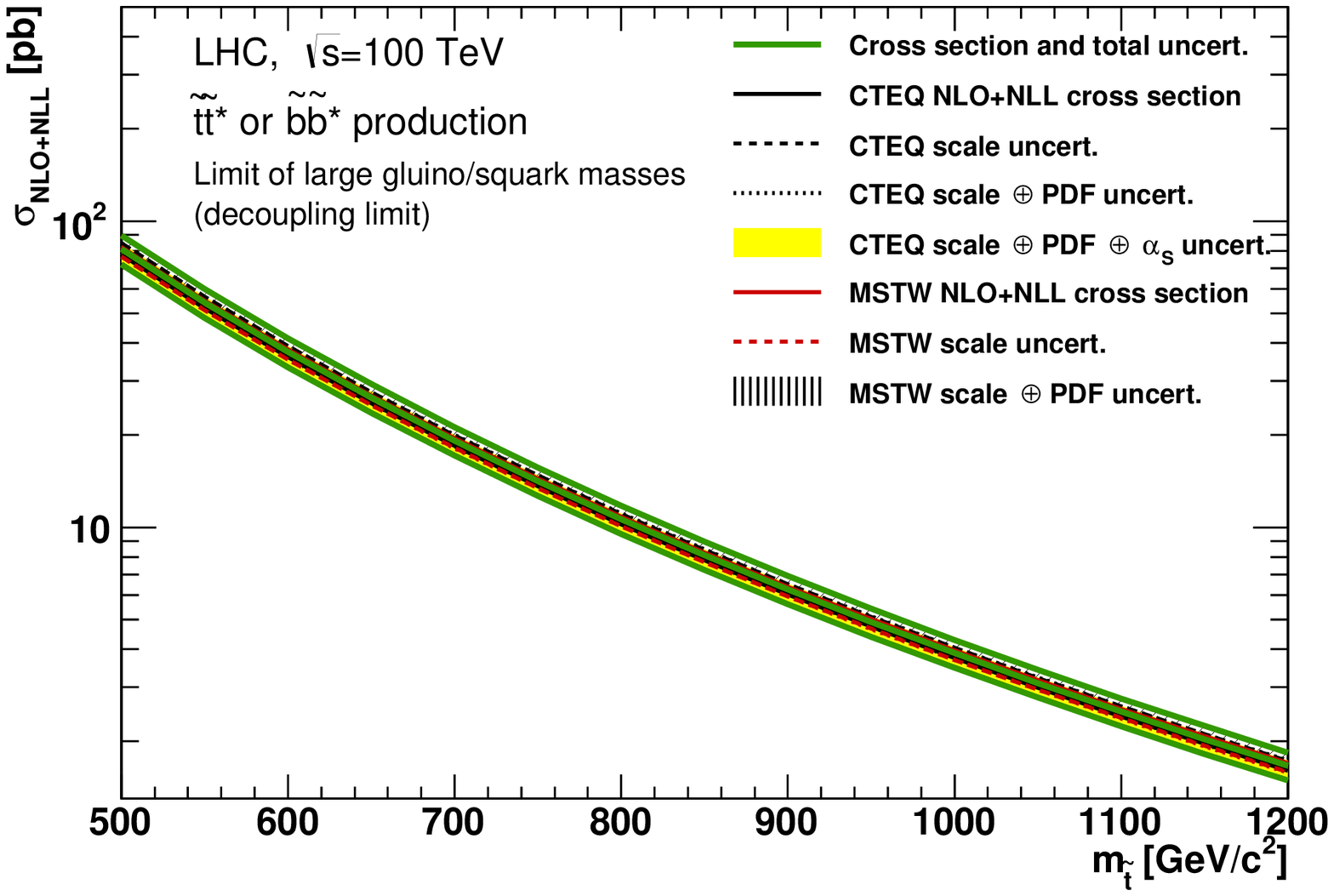}
  \end{center}
  \caption{NLO+NLL stop-antistop production cross section as a function of mass at $\sqrt{s}=100$ TeV in the wider (upper plot) and narrower (lower plot) mass range. The different 
   styled black (red) lines correspond to the cross section and scale uncertainties predicted
  using the \cteq\ (\mstw) PDF set. The yellow (dashed black) band
  corresponds to the total \cteq\ (\mstw) uncertainty, as described in
  the text. The green lines show the final cross section and its total
  uncertainty.}
  \label{fig:xsect_stop100}
\end{figure}

The production cross section as a function of the stop mass for a
model in which only the lightest stop is reachable is shown
in Figures~\ref{fig:xsect_stop13},~\ref{fig:xsect_stop14},~\ref{fig:xsect_stop33} and~\ref{fig:xsect_stop100} .  It should be noted that these cross
sections are approximately the same as those of a model in which only the lightest
sbottom is accessible, assuming the rest of the coloured SUSY spectrum
decoupled.

%% file: summary.tex

\section{Summary and future prospects}
\label{sec:summary}
We have presented reference cross sections for the production of squarks and gluinos at the upcoming LHC runs with $\sqrt{s}=13$ and $14$, and at future $pp$ colliders operating at $\sqrt{s}=33$ and $100$~TeV. The theoretical predictions are based on resummed results at the next-to-leading logarithmic (NLL) accuracy matched to next-to-leading order (NLO) predictions.\footnote{This paper should be cited along with
  the original papers, i.e.\ Refs.~\cite{Beenakker:1996ch, Kulesza:2008jb, Kulesza:2009kq, Beenakker:2009ha, Beenakker:2011fu}
 for inclusive squark/gluino production, Refs.~\cite{Beenakker:1997ut, Beenakker:2010nq, Beenakker:2011fu} 
for stop or sbottom direct production.}. We provide an estimate of the theoretical uncertainty following the prescriptions established in~\cite{Kramer:2012bx}, and used by the
\atlas\ and \cms\ collaborations in the interpretation of their measurements for $\sqrt{s}=7$ TeV and $8$ TeV.
The theoretical systematic uncertainties are
larger for higher sparticle masses, and they are typically dominated by the PDF
uncertainties. These have a significant impact when assessing the
experimental constraints or the sensitivity to a given SUSY model. 
Cross sections are evaluated using the~\cteq~and~\mstw~PDFs.
The large-x behavior of these PDF sets is determined in terms of few parameters, whose values are fixed 
in the region with experimental constraints. For the production of high-mass
SUSY particles, these functional forms are extrapolated beyond the constraints provided by data.
Differences between PDF sets will be reduced as more and more experimental 
measurements become available, in particular with the results of the LHC Run II, as well
as by improving the fitting methodology and the theoretical calculations.
In anticipation of this improved accuracy expected in future PDF
determinations, the central values presented for the first time in 
this paper at NLL accuracy can serve as an estimate of high mass SUSY coloured production for current and future 
colliders.  Detailed numbers and tables for a broad class of SUSY models 
and parameters are collected at the SUSY cross section working group web page~\cite{combined7TeV}.